\documentclass[preprint,preprintnumbers,amsmath,amssymb]{revtex4}
\usepackage{fancyhdr}
\usepackage{amssymb}
\usepackage{amsmath}
\usepackage[mathscr]{eucal}
\usepackage{latexsym}
\usepackage{amsbsy}
\usepackage{float}
\usepackage[dvips]{graphicx}
\usepackage{epsfig}
\usepackage{subfigure}
\usepackage{wrapfig}
\usepackage{epsf}
\usepackage{float}

\newcommand{\be}{\begin{equation}}
\newcommand{\ee}{\end{equation}}
\newcommand{\bes}{\begin{equation*}}
\newcommand{\ees}{\end{equation*}}
\newcommand{\bea}{\begin{eqnarray}}
\newcommand{\eea}{\end{eqnarray}}
\newcommand{\bean}{\begin{eqnarray*}}
\newcommand{\eean}{\end{eqnarray*}}
\newcommand{\ba}{\begin{array}}
\newcommand{\ea}{\end{array}}

\begin{document}
\title{\large Analytical approach to the two-site Bose-Hubbard model: 
from Fock states to Schr\"odinger cat states and entanglement entropy} 
\author{Luca Dell'Anna}
\affiliation{Dipartimento di Fisica e Astronomia ``Galileo Galilei'', Universit\`a di Padova, via F. Marzolo 8, 35131 Padova, Italy}
\begin{abstract}
We study the interpolation from occupation number Fock states to 
Schr\"odinger cat states 
on systems modeled by two-mode Bose-Hubbard Hamiltonian, 
like, for instance, bosons in a double well or superconducting 
Cooper pair boxes.
In the repulsive interaction regime, by a simplified single particle 
description, we calculate analytically energy, number fluctuations,  
stability under coupling to a heat bath, entanglement entropy and  
Fisher information, all in terms of 
hypergeometric polynomials of the single particle overlap parameter. 
Our approach allows us to find how those quantities scale with the 
number of bosons. 
In the attractive interaction regime 
we calculate the same physical quantities in terms 
of the imbalance parameter, and find that the spontaneous 
symmetry breaking, occurring at interaction $U_c$, 
predicted by a semiclassical approximation, is valid only in 
the limit of infinite number of bosons. 
For a large but finite number, we determine a characteristic 
strength of interaction, $U_c^*$, which can be promoted as the crossover 
point from coherent to incoherent regimes and can be identified as the 
threshold of fragility of the cat state. 
Moreover, we find that the Fisher information is always in direct ratio to the 
variance of on-site number of bosons, for both positive and 
negative interactions. 
We finally show that the entanglement entropy is maximum close to $U_c^*$ 
and exceeds its coherent value within the whole range of interaction between 
$2 U_c$ and zero.
\end{abstract}
\maketitle

\section{Introduction}
The two-mode Bose-Hubbard model is commonly used to describe several 
systems like, for 
examples, the bosonic double well traps \cite{imamoglu,castin,graham,cirac,steel,java} or the superconducting Cooper pair 
boxes \cite{jak,lew,abf,bfg,averin}. It has been recently used as the building 
block for studying spinful bosonic systems on double-well 
lattices \cite{bruder}. In spite of its simplicity, being an 
interacting problem, although integrable \cite{links} and whose 
low energy spectrum was explored in the weak interaction regime \cite{buonsante}, it is far from a simple solution for a generic number of bosons and 
interaction strength. 
Because of that reason, usually one resorts to numerical approaches which 
are simple and very efficient, giving up trying to describe the system 
analytically, although not exactly. A first issue worth being addressed is, 
therefore, that of providing a simple description which allows one to 
find handy and analytical expressions for several physical quantities 
and their scalings with the number of bosons.  
In order to discriminate between the so-called 
quantum phase model \cite{anglin}, expressed by occupation number states, 
and the mean field model \cite{dalfovo}, formulated in terms of coherent 
states, emerging from the solution of 
the classical Gross-Pitaevski equation - both models invoked to 
describe, for instance, the charge oscillations in the superconducting Cooper 
pair boxes - 
it can be useful to study certain stability properties \cite{abf}. 
The two models, in fact, behave quite differently in the presence of noise 
induced by a weakly coupled external environment, therefore it is crucial 
to predict how the relaxation time scales with the number of bosons.

In this paper we propose, therefore, to tackle the problem of 
describing the ground state of a two-mode Bose-Hubbard model from an 
intuitive and physically 
transparent description, although approximate. 
Our proposal is based on the observation that the norm of a Fock state, 
which is the ground state for strongly repulsive interacting bosons,
$|\psi_F\rangle=(a_L^\dagger)^n(a_R^\dagger)^k|0\rangle\,,$
(where $a_{L,R}^\dagger$ are creation operators on left, $L$, or right, $R$, site) is given by the permanent (Per) of a diagonal block overlap matrix
\begin{displaymath}
\langle \psi_F|\psi_F\rangle=\textrm{Per}\left(
\ba{c|c}
1_{n\times n}&0_{n\times k}\\\hline
0_{k\times n}&1_{k\times k}
\ea
\right)\,,
\end{displaymath}
where, schematically, the blocks denoted by ``$1$'' are made of all elements 
equal to one and the blocks denoted by ``$0$'' are made of all null elements. 
On the other hand, the norm of a coherent-like state, or sometimes 
called phase state, ground state for free bosons,
$|\psi_C\rangle=(\xi_{L}a_L^\dagger+\xi_{R}a_R^\dagger)^{n+k}|0\rangle\,,$
with $|\xi_L|^2+|\xi_R|^2=1$, is given by the permanent of a fully 
constant matrix
\begin{displaymath}
\langle \psi_C|\psi_C\rangle=\textrm{Per}
\left(
\ba{c|c}
1_{n\times n}&1_{n\times k}\\\hline
1_{k\times n}&1_{k\times k}
\ea
\right)\,.
\end{displaymath}
A natural expectation is, therefore, that, going from number Fock state to 
fully delocalized coherent-like state, the bosons, initially localized on 
the two sites, start to overlap, making finite the off-diagonal blocks of 
the overlap matrix. 
Our ansatz is that the intermediate interaction regime can be fairly described 
by the state $|\psi\rangle$, written in Eq.~(\ref{wf}), that describes two 
condensates which can be localized on each site in the Fock limit while 
merging together in the coherent one, and whose norm is given by
\begin{displaymath}
\langle \psi|\psi\rangle=\textrm{Per}
\left(
\ba{c|c}
1_{n\times n}&\omega^*_{n\times k}\\\hline
\omega_{k\times n}&1_{k\times k}
\ea
\right)\,,
\end{displaymath}
where $\omega$ is indeed the overlap of the single particle wavefunctions, 
as we will be seeing. In the site-symmetric case, then, this description 
allows us to interpolate from Fock to coherent-like states by varying a single 
parameter, $\omega$, 
which has a clear physical meaning and can be fixed variationally in terms of 
the number of bosons and the microscopic parameters of the Hubbard model. 

Another important issue is related to the spontaneous symmetry breaking 
between the two wells, which is supposed to occur in the attractive 
interaction regime.
The experimental realization of a double-well potential which confines
ultracold alkali-metal atoms, is the atomic analog of the superconducting
Joshepson junction \cite{cataliotti, shin, albiez, levy}, as predicted several 
years ago \cite{java86}. 
The Josephson equations are valid in the weak interaction 
regime and predicts self-trapping and symmetry breaking of the atomic 
population in the two wells \cite{milburn,augusto,augusto2,toigo}. 
The symmetry breaking can be derived  starting from quasiclassical coherent 
states. We will consider, instead, coherent-like states, 
i.e. $|\psi_C\rangle$, given by the contributions to the full 
coherent states with fixed number of particles, which is the case in 
real experimental situations. Also these states exhibit a symmetry breaking, 
however we show that a symmetric linear combination of two of such states, 
namely a macroscopic Schr\"odinger-cat state \cite{cirac,dalvit,huang}, which 
has null population imbalance, is energetically favoured, for equal local 
energies. As a result, in the case of fixed number of particles, the symmetry 
breaking should not occur and the  
imbalance parameter, which does not correspond to the population 
imbalance, is finite for any attractive interaction. 
This result is in agreement also with numerical results \cite{mazzarella}, 
and, at some extent, consistent with the semiclassical approach in the 
limit $N\rightarrow \infty$.
We find, however, a discontinuity in the imbalance parameter at some critical 
strength of interaction. This is a spurious effect of our ansatz for the 
ground state, however it reveals the fragility of the cat state below a 
certain value of the interaction. Nevertheless, by numerical checks, we 
observed that, around that value of interaction, some physical quantities have 
large derivatives and the energy changes its behavior. At that point the 
coherence visibility drops while the cat state becomes extremely fragile. The 
relaxation time, in fact, goes like $1/N^2$ upon an asymmetric coupling with 
an external bath. One can identify such a threshold as the critical 
interaction below which an infinitesimally small mismatch between 
the two on-site energies 
can produce a macroscopic population imbalance. 

Finally we investigate the behavior of the entanglement entropy and the 
Fisher information. 
The quantum entanglement applied to many body systems has attracted a lot of 
theoretical interest in recent years (see, for instance, 
Refs. \cite{amico,haque} and references therein). One would expect 
that the entanglement entropy reaches its maximum in the presence of large 
coherence. We show, instead, that, in agreement with numerical results 
\cite{mazzarella}, the maximum entropy occurs close to the putative symmetry 
breaking point. In particular we found that the entropy exceeds its value 
obtained in the case of free bosons, in a whole range of attractive 
interaction which goes from 
zero to twice the critical interaction of the symmetry breaking. 
We finally show that, for almost all the range of interaction, i.e. 
$-\infty<U\ll N$, the Fisher information $F$ is always directly proportional 
to the variance of on-site number of bosons $\sigma$, as described by 
the equation  $F=4\sigma/N^2$. 

The paper is organized as follows. In Sec. II we introduce some useful 
definitions; in Sec. III we present the model and our ansatz 
for the approximate ground state in the repulsive interaction regime; 
in Sec. IV we perform detailed calculations for generic two and 
four-point correlation functions, useful to calculate several quantities of 
interest; in Sec. V and VI we recover the exact results, respectively, 
for strongly repulsive interacting and free bosons; 
in Sec. VII we derive asymptotic behaviors in the large $N$ limit 
for energy, decay rate, number fluctuations and coherent visibility; 
in Sec. VIII we calculate, by our ansatz for repulsive interaction 
ground state, the entanglement entropy and the Fisher information.
Section IX, instead, is devoted to the analysis of attractive interaction, 
calculating all the quantities of interest as Schr\"odinger cat state 
mean values and finally in Sec. X we derive entanglement entropy and 
Fisher information, always in the attractive regime. 
We summarize the main results showing some plots in Sec. XI, and 
drawing some conclusions in the final section. 
\section{General remarks}
Let us first consider $N$ bosons on a lattice with $N_s$ sites and take 
the following, not normalized, many-body wavefunctions
\bea
\label{psiN}
|\psi\rangle&=&\prod_{\alpha=1}^{N}\left(\sum_{i=1}^{N_s}\xi_{\alpha i}\, a^\dagger_{i}\right)|0\rangle\,,\\
|\varphi\rangle&=&\prod_{\alpha=1}^{N}\left(\sum_{i=1}^{N_s}\eta_{\alpha i}\, a^\dagger_{i}\right)|0\rangle\,.
\eea
where $\xi_{\alpha i}$ and $\eta_{\alpha i}$ are single particle wavefunctions.
It has been shown that such states are rich enough to exhibit both 
superfluid and insulating behaviors \cite{jstat}. 
After defining the $N\times N$ matrix
\be
\Omega_{\alpha \beta}=\sum_{i}  \eta_{\alpha i}\xi^{*}_{\beta i} \;,
\ee
the $(N+1)\times (N+1)$ matrix
\be
D_{ij}=
\left(
\ba{cc}
\hat\Omega &\hat\eta_{i}\\
\hat\xi^\dagger_j& \delta_{ij}
\ea
\right) \,,
\ee
and the $(N+2)\times (N+2)$ matrix
\be
I_{lijm}=
\left(
\ba{ccc}
\hat\Omega &\hat\eta_{i}& \hat\eta_{l}\\
\hat\xi^\dagger_j& \delta_{ij}& \delta_{il}\\
\hat\xi^\dagger_m& \delta_{im}& \delta_{ml}
\ea
\right)\,,
\ee
where $\hat\eta_{i}$ is an $N$-vector with components $\eta_{\alpha i}$, 
we get the following relations \cite{jstat}
\bea
\langle \psi|\varphi\rangle &=& \textrm{Per}(\Omega)\,,\\
\langle \psi|\,a_l\, a^{\dagger}_m|\varphi\rangle &=& \textrm{Per}(D_{lm})\,,\\
\langle \psi|\,a_l\,a_i\,a^{\dagger}_j\,a^{\dagger}_m|\varphi\rangle &=& \textrm{Per}(I_{lijm})\,,
\eea
where Per($A$) is the permanent of the matrix $A$. Quite in general, since 
the calculus of permanents is not easy, it is hard to find handy and 
analytical expressions for the correlation functions. 

\section{The interpolating state}
Now let us take $N=n+k$ bosons on two sites, like in a double well, 
and consider the following many-body state  
\be
\label{wf}
|\psi\rangle=\left(\xi_{1L}a_L^\dagger+\xi_{1 R}a_R^\dagger\right)^n\left(\xi_{2 L}a_L^\dagger+\xi_{2R}a_R^\dagger\right)^{k}|0\rangle
\ee
which is a simplified version of the permanent in Eq.~(\ref{psiN}), 
where $\forall \alpha$ such that $\alpha\le n$, $\xi_{\alpha i}=\xi_{1i}$ while $\forall \alpha$ such that $n<\alpha\le (n+k)$, $\xi_{\alpha i}=\xi_{2i}$. 
Defining the unit $n\times k$ matrix (with $n$ rows and $k$ columns) 
with all elements equal to $1$,
\be
J^{(n,k)}=
\left(
\ba{cccc}
1&1&\dots&1\\
:&:&\dots&:\\
1&1&\dots&1
\ea
\right)\,,
\ee
we can construct the overlap matrix 
\be
\Omega^{n,k}=\left(
\ba{cc}
b J^{(n,n)}&\omega^* J^{(n,k)}\\
\omega J^{(k,n)}&b J^{(k,k)}
\ea\right)\,,
\ee
where 
\be
\label{normaliz}
b=\left|\xi_{1L}\right|^2+\left|\xi_{1R}\right|^2=\left|\xi_{2L}\right|^2+\left|\xi_{2R}\right|^2
\ee
is the normalization of the single particle wavefunction, therefore we can set 
$b=1$. However, for the moment, in order to derive more general relations, 
we will keep the writing $b$. The other parameter is
\be
\label{overl}
\omega=\xi_{2L}\xi_{1L}^*+\xi_{2R}\xi_{1R}^*
\ee
which is, therefore, the single particle overlap term. For normalized single particle wavefunctions, namely when $b=1$, we have $0\le |\omega|\le 1$.

Eq.~(\ref{wf}) is the simplest permanent state which interpolates, 
keeping the same functional form, from number Fock state (when, for instance, 
$\xi_{1L}=\xi_{2R}=1$ and $\xi_{2L}=\xi_{1R}=0$, such that we have $n$ bosons 
on the left well and $k$ bosons on the right well) to a coherent-like state 
(when $\xi_{1L}=\xi_{2L}$ and $\xi_{1R}=\xi_{2R}$).

The four parameters, $\xi_{1L}, \xi_{2L}, \xi_{1R}, \xi_{2R}$, are not 
independent, but linked by the two normalization conditions, 
Eq.~(\ref{normaliz}). 
On the contrary Eq.~(\ref{overl}) is not a further constraint but leads to a 
different parametrization, where the overlap coefficient $\omega$ replaces one of the $\xi$-parameters. Moreover, from Eq.~(\ref{wf}), one can consider $n$ as 
another free parameter while $k$ is fixed by $k=N-n$.
Therefore, at the end, we have three free state parameters (in general two 
complex numbers and one integer): $\xi_{1L}$ (one among the four $\xi$'s), 
the overlap $\omega$ and $n$. 

These values can be related to the parameters of the microscopic model. 
Let us introduce the Bose-Hubbard model
\be
\label{BH}
H=-\frac{t}{2}\left(a_L^\dagger a_R+a_R^\dagger a_L\right)+\mu\left(n_L-n_R\right)+\frac{U}{2}\big(n_L(n_L-1)+n_R(n_R-1)\big)\,,
\ee
where $t$ is the hopping parameter, $\mu$ is the difference of local 
chemical potentials induced for instance by an mismatch 
between the two wells, and 
$U$ is the on site interaction. One can therefore link the three parameters 
$\xi_{1L}, \omega, n$ with the three parameters of the Bose-Hubbard model, 
$t, \mu, U$, by variational analysis. Actually, defining
\be
E\left[\xi_{1L}, n, \omega | t, \mu, U\right]=\frac{\langle\psi|H|\psi\rangle}{\langle\psi|\psi\rangle}\,,
\ee 
one can partially fix the set of state parameters, requiring that 
$\delta_{\xi_{1L}}E=
\delta_{\omega}E=
\delta_{n}E=0$, 
establishing, in this way, a connection between the set of the state parameters with the set of the Hamiltonian ones. The residual arbitrariness can be 
removed if we require that the amplitudes $\xi_{\alpha i}$ are real. 

\section{Correlation functions}
In order to calculate two and four points correlation functions we have to 
determine the permanents of the matrices $\Omega, D$ and $I$. As we will see, 
all these permanents can be calculated analytically.  
Let us start considering 
the normalization of the many-body wavefunction Eq.~(\ref{wf}). 
This quantity is given by 
\be
\label{PerOmega}
\langle\psi|\psi\rangle=\textrm{Per}(\Omega^{n,k})=n! k! \,b^{n+k} \phantom{.}_2F_1\left(-n,-k,1;\frac{|\omega|^2}{b^2}\right)
\ee
which is derived by induction, applying the Laplace theorem for the permanent and where 
$_2F_1\left(a,b,1;z\right)$ is an hypergeometric function. 
Since the first arguments are integer numbers, $_2F_1$ is a polynomial, 
particularly, using the definition of the Jacobi polynomials 
$P^{(\alpha,\beta)}_n(z)$, it can be written as follows
\be
\phantom{.}_2F_1\left(-n,-k,1;z\right)=P^{(0,-n-k-1)}_n(1-2z)\,.
\ee
Actually, one can verify that $\textrm{Per}(\Omega^{n,k})$ satisfies the 
following recursive equation
\be
\textrm{Per}(\Omega^{n,k})=n\,b\textrm{Per}(\Omega^{n-1,k})+k!\,|\omega|^2 b^{k-2}
\left[n!\,b^{n} +n\sum_{\ell=1}^{k-1}\frac{b^{1-\ell}}{\ell!}
\textrm{Per}(\Omega^{n-1,\ell})\right]
\ee
and check that Eq.~(\ref{PerOmega}) is a solution.
In order to calculate two and four point correlation functions it is convenient to calculate the permanents of the following matrices
\bea
{\cal O}^{n,k}_{1u}&=&\left(
\ba{ll}
b J^{(n-1,n)}&\omega^* J^{(n-1,k-1)}\\
\omega J^{(k,n)}&b J^{(k,k-1)}
\ea\right)\,,\\
{\cal O}^{n,k}_{1d}&=&\left(
\ba{ll}
b J^{(n,n-1)}&\omega^* J^{(n,k)}\\
\omega J^{(k-1,n-1)}&b J^{(k-1,k)}
\ea\right)\,,
\eea
with dimensions $(n+k-1)\times (n+k-1)$ and
\bea
{\cal O}^{n,k}_{2u}&=&\left(
\ba{ll}
b J^{(n-2,n)}&\omega^* J^{(n-2,k-2)}\\
\omega J^{(k,n)}&b J^{(k,k-2)}
\ea\right)\,,\\
{\cal O}^{n,k}_{2d}&=&\left(
\ba{ll}
b J^{(n,n-2)}&\omega^* J^{(n,k)}\\
\omega J^{(k-2,n-2)}&b J^{(k-2,k)}
\ea\right)\,,
\eea
with dimensions $(n+k-2)\times (n+k-2)$. 
Their permanents, together with Eq.~(\ref{PerOmega}), are the building blocks 
useful to construct the permanents of the $D$-matrix, for the two-point correlation functions, and of the $I$-matrix, for the four-point correlation 
functions. By recursion, after several algebraic steps, we get
\bea
\label{perO1}
&&\textrm{Per}({\cal O}^{n,k}_{1u})=n! (k-1)!\, \omega\,b^{n+k-2}\sum_{\ell=0}^{k-1}\phantom{.}_2F_1\left(1-n,-\ell,1;\frac{|\omega|^2}{b^2}\right),\\
&&\textrm{Per}({\cal O}^{n,k}_{1d})=(n-1)!k!\, \omega^*\,b^{n+k-2}\sum_{\ell=0}^{n-1}\phantom{.}_2F_1\left(-\ell,1-k,1;\frac{|\omega|^2}{b^2}\right),\;\\
&&\textrm{Per}({\cal O}^{n,k}_{2u})=k (n-2)!(k-2)!\, \omega^2\,b^{n+k-4}
\sum_{m=0}^{n-2}\left[(1+m)\sum_{\ell=0}^{k-2}
\phantom{.}_2F_1\left(-m,-\ell,1;\frac{|\omega|^2}{b^2}\right)\right],\;\\
&&\textrm{Per}({\cal O}^{n,k}_{2d})=n (n-2)!(k-2)!\, \omega^{*2}\,b^{n+k-4}\sum_{\ell=0}^{k-2}
\left[(1+\ell)\sum_{m=0}^{n-2}
\phantom{.}_2F_1\left(-m,-\ell,1;\frac{|\omega|^2}{b^2}\right)\right].\,
\label{perO2}
\eea
However we will need only a couple of those permanents since
\bea
&&\textrm{Per}({\cal O}^{n,k}_{1u})=\textrm{Per}({\cal O}^{n,k}_{1d})^*\,,\\
&&\textrm{Per}({\cal O}^{n,k}_{2u})=\textrm{Per}({\cal O}^{n,k}_{2d})^*\,.
\eea
We now write the $(n+k+1)\times (n+k+1)$-matrix 
\be
D^{n,k}_{ij}=\left(
\ba{ccc}
b J^{(n,n)}&\omega^* J^{(n,k)}&\hat\xi_{1i}\\
\omega J^{(k,n)}&b J^{(k,k)}&\hat\xi_{2i}\\
\hat\xi^\dagger_{1j}&\hat\xi^\dagger_{2j}&\delta_{ij}
\ea\right)\,,
\ee
where $i,j=L,R$ and 
with $\hat\xi^\dagger_{1j}=\xi^*_{1j}(1,1,...,1)$, a $n$-vector and $\hat\xi^\dagger_{2j}=\xi^*_{2j}(1,1,...,1)$, a $k$-vector.
Its permanent, related to two-point correlation functions, can be written as follows
\bea
\langle\psi|a_ia_j^\dagger|\psi\rangle=\nonumber\textrm{Per}(D^{n,k}_{ij})=\delta_{ij} \textrm{Per}(\Omega^{n,k})
+n\xi_{1i}\left[n\xi^*_{1j}\textrm{Per}(\Omega^{n-1,k})
+k\xi^*_{2j}\textrm{Per}({\cal O}^{n,k}_{1u})\right]\\
+k\xi_{2i}\left[k\xi^*_{2j}\textrm{Per}(\Omega^{n,k-1})+n\xi^*_{1j}\textrm{Per}({\cal O}^{n,k}_{1d})\right]\,.
\label{PerD}
\eea
Now defining the $(n+k+2)\times (n+k+2)$ matrix
\be
I^{n,k}_{lijm}=\left(
\ba{cccc}
b J^{(n,n)}&\omega^* J^{(n,k)}&\hat\xi_{1i}&\hat\xi_{1l}\\
\omega J^{(k,n)}&b J^{(k,k)}&\hat\xi_{2i}&\hat\xi_{2l}\\
\hat\xi^\dagger_{1j}&\hat\xi^\dagger_{2j}&\delta_{ij}&\delta_{lj}\\
\hat\xi^\dagger_{1m}&\hat\xi^\dagger_{2m}&\delta_{im}&\delta_{lm}
\ea\right)\,,
\ee
with $l,i,j,m=L,R$
we get
\bea
\nonumber\langle\psi|a_la_ia_j^\dagger a_m^\dagger|\psi\rangle&=&\textrm{Per}(I^{n,k}_{lijm})\\
&&\hspace{-2.0cm}\nonumber=\delta_{lm}\textrm{Per}(D^{n,k}_{ij})+\delta_{lj}
\textrm{Per}(D^{n,k}_{im}) +\delta_{im}\textrm{Per}(D^{n,k}_{lj})-\delta_{lj}\delta_{im}\textrm{Per}(\Omega^{n,k})\\
&&\hspace{-2.0cm}\nonumber +k\xi_{2l}\left\{
n\xi^*_{1m}\left[\delta_{ij}\textrm{Per}({\cal O}^{n,k}_{1d})+n\xi_{1i}
\left(k\xi^*_{2j}\textrm{Per}(\Omega^{n-1,k-1})+(n-1)\xi^*_{1j}
\textrm{Per}({\cal O}^{n-1,k}_{1d})\right)\right.\right.\\
&&\hspace{-2.0cm}\nonumber \left.\left.+(k-1)\xi_{2i}\left((n-1)\xi^*_{1j}\textrm{Per}({\cal O}^{n,k}_{2d})+k\xi^*_{2j}\textrm{Per}({\cal O}^{n,k-1}_{1d})\right)\right]
+k\xi^*_{2m}\textrm{Per}(D^{n,k-1}_{ij})
\right\}\\
&&\hspace{-2.0cm}\nonumber +n\xi_{1l}\left\{
k\xi^*_{2m}\left[\delta_{ij}\textrm{Per}({\cal O}^{n,k}_{1u})+k\xi_{2i}
\left(n\xi^*_{1j}\textrm{Per}(\Omega^{n-1,k-1})+(k-1)\xi^*_{2j}\textrm{Per}({\cal O}^{n,k-1}_{1u})\right)\right.\right.\\
&&\hspace{-2.0cm}\left.\left.+(n-1)\xi_{1i}\left((k-1)\xi^*_{2j}\textrm{Per}({\cal O}^{n,k}_{2u})+n\xi^*_{1j}\textrm{Per}({\cal O}^{n-1,k}_{1u})\right)\right]
+n\xi^*_{1m}\textrm{Per}(D^{n-1,k}_{ij})
\right\}\,.
\label{PerI}
\eea
Now we are in the position to calculate analytically, in terms of the Jacobi 
polynomials, several quantities, like the total energy, the number 
fluctuations, the decay rate when the system is coupled to an external 
environment and the coherence visibility.
\subsection{Energy}
Supposing that our system is described by the Hamiltonian given in Eq.~(\ref{BH}), we can calculate the energy of our many body state, Eq.~(\ref{wf}), as 
follows
\bea
\label{totalE}
\nonumber E=\langle H\rangle &=&-\frac{t}{2}\left(
\frac{\textrm{Per}(D^{n,k}_{LR})+\textrm{Per}(D^{n,k}_{RL})}
{\textrm{Per}(\Omega^{n,k})}\right)
+\frac{U}{2}\left(
\frac{\textrm{Per}(I^{n,k}_{LLLL})+\textrm{Per}(I^{n,k}_{RRRR})}
{\textrm{Per}(\Omega^{n,k})}\right)\\
&&+(\mu-2U)\frac{\textrm{Per}(D^{n,k}_{LL})}{\textrm{Per}(\Omega^{n,k})}-
(\mu+2U)\frac{\textrm{Per}(D^{n,k}_{RR})}{\textrm{Per}(\Omega^{n,k})}+2U \;.
\eea

\subsection{Number fluctuations}
In the same manner we can calculate the charge fluctuations, for example on the left site, 
\be
\label{sigmaL}
\sigma_L=\langle n_L^2 \rangle-\langle n_L \rangle^2\,,
\ee
as the variance of the number operator, where 
\bea
\langle n_L^2\rangle&=&\frac{\langle\psi| a^\dagger_{L}a_{L}a^\dagger_{L}a_{L}|\psi\rangle}
{\langle\psi|\psi\rangle}=
\frac{\textrm{Per}(I^{n,k}_{LLLL})-3\textrm{Per}(D^{n,k}_{LL})}
{\textrm{Per}(\Omega^{n,k})}+1\,,\\
\langle n_L\rangle^2&=&\left(\frac{\langle\psi|a^\dagger_{L}a_{L}|
\psi\rangle}{\langle\psi|\psi\rangle}\right)^2=
\left(\frac{\textrm{Per}(D^{n,k}_{LL})}{\textrm{Per}(\Omega^{n,k})}-1\right)^2\,.
\eea
Analogously one can write the charge fluctuations on the right site, 
namely $\sigma_R$.

\subsection{Decay rate}
In this subsection we suppose our system weakly coupled to an external 
environment, therefore the total Hamiltonian is \cite{fabio}
\be
H_T=H+H_B+\lambda \left({W} \hat{B}+{W}^\dagger \hat{B}^\dagger\right)\,,
\ee
where $H_B$ is the environment Hamiltonian while the last term is the weak coupling between the bath and the bosons, i.e. $\lambda\ll 1$. ${W}$ is a bosonic operator. The heat bath acts as a source of noise and dissipation. We assume the environment to be in an equilibrium state and that behaves like a white noise, $\langle\hat{B}(t)^\dagger\hat{B}\rangle_E
\simeq \delta(t)$. The presence of the bath leads to a master equation for 
the density matrix, $\rho=\frac{|\psi\rangle\langle\psi|}
{\langle\psi|\psi\rangle}$, of the Kossakowski-Lindblad form \cite{gorini}
\be
\partial_t\rho=-i[H+H^{(2)},\rho]+{\cal{D}}[\rho]\,,
\ee
where
\bea
\nonumber {\cal{D}[\rho]}=\gamma\left({W}\rho {W}^\dagger-\frac{1}{2}\{{W}^\dagger{W},\rho\}\right)
+\delta\left({W}^\dagger\rho {W}-\frac{1}{2}\{{W}{W}^\dagger,\rho\}\right)\\
+\beta\left({W}\rho {W}-\frac{1}{2}\{{W}^2,\rho\}\right)
+\beta^*\left({W}^\dagger\rho {W}^\dagger-\frac{1}{2}\{{W}^{\dagger 2},\rho\}\right)\,,
\eea
with $\gamma=\lambda^2\int^{+\infty}_{-\infty}dx \langle \hat{B}^\dagger(x)\hat{B}\rangle_E$,
$\,\delta=\lambda^2\int^{+\infty}_{-\infty}dx \langle \hat{B}(x)\hat{B}^\dagger\rangle_E$,
and $\beta=\lambda^2\int^{+\infty}_{-\infty}dx \langle \hat{B}(x)\hat{B}\rangle_E$, some coefficients, $\beta\in \mathbb{C}$, $\gamma\ge 0$, $\delta\ge 0$, 
satisfying the condition $\gamma\delta\ge |\beta|^2$, which ensures the 
complete positivity (see Ref. \cite{fabio} for more details). 
These coefficients incorporate the dissipative 
effects on the dynamics. The term $H^{(2)}$, 
is an environment induced additional Hamiltonian term whose explicit expression is not important for our purposes.
Here, in fact, we focus our attention to the stability of our initial pure state $\frac{|\psi\rangle\langle\psi|}{\langle\psi|\psi\rangle}$ at $t=0$, calculating the constant decay rate 
defined by
\be
\Gamma=-\frac{\langle\psi|\partial_t\rho |\psi\rangle}{\langle\psi|\psi\rangle}
\Big|_{t=0}=-\frac{\langle \psi|{\cal{D}}\big[|\psi\rangle\langle\psi|\big]
|\psi\rangle}{\langle\psi|\psi\rangle^2}= \gamma \,\Gamma_\gamma+\delta\, 
\Gamma_\delta+ 2 \textrm{Re}[\beta \,\Gamma_\beta]\,,
\ee
where
%
\bea 
\label{decay}
&&\Gamma_\gamma=\frac{\langle\psi|{W}^\dagger{W}|\psi\rangle}{\langle \psi|\psi\rangle}-
\left|\frac{\langle\psi|{W}|\psi\rangle}{\langle\psi|\psi\rangle}\right|^2\\
&&\Gamma_\delta=\frac{\langle\psi|{W}{W}^\dagger|\psi\rangle}{\langle \psi|\psi\rangle}-
\left|\frac{\langle\psi|{W}|\psi\rangle}{\langle\psi|\psi\rangle}\right|^2\\
&&\Gamma_\beta= \frac{\langle\psi|{W}{W}|\psi\rangle}{\langle \psi|\psi\rangle}-\left(\frac{\langle\psi|{W}|\psi\rangle}{\langle\psi|\psi\rangle}\right)^2
\eea
Let us choose ${W}$ as a generic single particle operator, linear combination of an hopping term and right and left density operators
\be
{W}=c_L\, n_L +c_R\, n_R + c_h \,a_R^\dagger a_L\,.
\ee
Usually only the hopping operator is considered in the coupling with the bath 
\cite{abf}, i.e. $c_R=c_L=0$. Here, instead, we consider a more general 
operator.
Without loss of generality and without spoiling the complete positivity, we 
could put $\beta=0$ and consider only $\Gamma_\gamma$ or $\Gamma_\delta$ 
to show how the decay rate scales with the number of bosons.
We have therefore
\bea
\nonumber\Gamma_\gamma&=&|c_L|^2\,\sigma_L+|c_R|^2\,\sigma_R+
|c_h|^2\,\Gamma^h_\gamma+2\textrm{Re}(c_Lc_R^*)\,
\big(\langle n_Ln_R\rangle-\langle n_L\rangle\langle n_R\rangle\big)\\
\nonumber&+&c_Lc_h^*\big(\langle n_La_L^\dagger a_R\rangle-\langle n_L\rangle\langle a_L^\dagger a_R\rangle\big)+c_hc_L^*\big(\langle a_R^\dagger a_L n_L\rangle-\langle n_L\rangle\langle a_R^\dagger a_L\rangle\big)\\
&+&c_Rc_h^*\big(\langle n_Ra_L^\dagger a_R\rangle-\langle n_R\rangle\langle a_L^\dagger a_R\rangle\big)+c_hc_R^*\big(\langle a_R^\dagger a_L n_R\rangle-\langle n_R\rangle\langle a_R^\dagger a_L\rangle\big)
\eea
where $\langle ...\rangle\equiv\frac{\langle\psi|...|\psi\rangle}{\langle\psi|\psi\rangle}$, $\sigma_L$ and $\sigma_R$ are defined by Eq.~(\ref{sigmaL}) and 
\be
\label{decayh}
\Gamma^h_\gamma=                                           
{\langle a^\dagger_{L}a_{R}a^\dagger_{R}a_{L}\rangle}
-\big|{\langle a_{L} a^\dagger_{R}\rangle}
\big|^2                                             
=\frac{\textrm{Per}(I^{n,k}_{LRRL})-\textrm{Per}(D^{n,k}_{LL})}{\textrm{Per}
(\Omega^{n,k})}-                   
\left|\frac{\textrm{Per}(D^{n,k}_{LR})}{\textrm{Per}(\Omega^{n,k})}\right|^2
\ee
which is the common $\gamma$-contribution to the decay rate when the bath induces only hopping of particle between the two sites \cite{abf}. When the environment couples also to the local densities we have to calculate, in addition, 
\bea
&&\hspace{-0.8cm}\langle n_Ln_R\rangle-\langle n_L\rangle\langle n_R\rangle=\frac{\textrm{Per}(I^{n,k}_{LRLR})}{\textrm{Per}(\Omega^{n,k})}-\frac{\textrm{Per}(D^{n,k}_{LL})
\textrm{Per}(D^{n,k}_{RR})}{\textrm{Per}(\Omega^{n,k})^2}\,,\\
&&\hspace{-0.8cm}\langle n_La_L^\dagger a_R\rangle-\langle n_L\rangle
\langle a_L^\dagger a_R\rangle=\frac{\textrm{Per}(I^{n,k}_{LRLL})}{\textrm{Per}(\Omega^{n,k})}-\frac{\textrm{Per}(D^{n,k}_{LR})}{\textrm{Per}(\Omega^{n,k})}-
\left(\frac{\textrm{Per}(D^{n,k}_{LL})}{\textrm{Per}(\Omega^{n,k})}-1\right)
\frac{\textrm{Per}(D^{n,k}_{RL})}{\textrm{Per}(\Omega^{n,k})}\,,\\
&&\hspace{-0.8cm}\langle n_Ra_L^\dagger a_R\rangle-\langle n_R\rangle
\langle a_L^\dagger a_R\rangle=\frac{\textrm{Per}(I^{n,k}_{RRLR})}
{\textrm{Per}(\Omega^{n,k})}-2
\frac{\textrm{Per}(D^{n,k}_{LR})}{\textrm{Per}(\Omega^{n,k})}-
\left(\frac{\textrm{Per}(D^{n,k}_{RR})}{\textrm{Per}(\Omega^{n,k})}-1\right)
\frac{\textrm{Per}(D^{n,k}_{RL})}{\textrm{Per}(\Omega^{n,k})}\,,
\eea 
and conjugate terms. Analogous calculations can be done for $\Gamma_\delta$ 
and $\Gamma_\beta$.

\subsection{Visibility}
In cold atom physics, in order to detect the coherence
properties of the condensates, the quantity which are commonly used is
the momentum distribution, namely the Fourier transform of the one-body density
matrix $C(x,x')=\langle a(x)^\dagger a(x')\rangle$, where the brackets
mean the ground state average. It has been shown \cite{stringari,stringari2,anna} 
that the
momentum distribution, $n(p)=\int dx dx' e^{-ip(x-x')}C(x,x')$, can be 
written as $n(p)=n_0(p)(1+\alpha \cos(p d))$, where $n_0(p)$ is the 
momentum distribution in the incoherent regime which depends on the details 
of the feasible double-well potential, $d$ the distance of the two minima 
of the double-well and $\alpha$ the so-called visibility. 
In terms of our site-operators, the visibility can be defined by
\be
\alpha=\frac{2|\langle a_L^\dagger a_R\rangle|}{N}\,,
\ee
which, if valuated using our permanent states, is simply given by 
\be
\label{vis2}
\alpha=\frac{2}{N}\left|\frac{\textrm{Per}(D^{n,k}_{RL})}
{\textrm{Per}(\Omega^{n,k})}\right| .
\ee

\section{Fock limit, $\omega=0$}
Here and in what follows we put $b=1$, namely we fix the normalization of the 
single particle wavefunctions. 
For $\omega=0$ we have $\textrm{Per}({\cal O}_{1u})=\textrm{Per}({\cal O}_{1d})=\textrm{Per}({\cal O}_{2u})=\textrm{Per}({\cal O}_{d2})=0$ while $\phantom{.}_2F_1\left(-n,-k,1;0\right)=1$, $\forall n, k$, therefore
\bea
&&\textrm{Per}(\Omega^{n,k})=n! k! \;,\\
\label{PerDFock}
&&\textrm{Per}(D^{n,k}_{ij})=n! k! \left(\delta_{ij}+n \xi_{1j}^*\xi_{1i}
+k\xi_{2j}^* \xi_{2i}\right)\;,\\
\label{PerIFock}
&&\nonumber\textrm{Per}(I^{n,k}_{lijm})=n! k! \left[\delta_{ij}\delta_{lm}+\delta_{im}\delta_{lj}+
n\left(\delta_{lm}\xi_{2j}^* \xi_{2i}+\delta_{im}\xi_{2j}^* \xi_{2l}+\delta_{lj}\xi_{2m}^* \xi_{2i}+\delta_{ij}\xi_{2m}^* \xi_{2l}\right)  \right.\\
&&\nonumber\phantom{--} +k\left(\delta_{lm}\xi_{1j}^* \xi_{1i}+
\delta_{im}\xi_{1j}^* \xi_{1l}+
\delta_{lj}\xi_{1m}^* \xi_{1i}+\delta_{ij}\xi_{1m}^* \xi_{1l}\right)+
(n^2-n)\xi_{1i}\xi_{1j}^* \xi_{1l}\xi_{1m}^*\\
&&\phantom{--}\left.+(k^2-k)\xi_{2i}\xi_{2j}^*\xi_{2l}\xi_{2m}^*
+nk\Big(\xi_{1l}\xi_{2i}+\xi_{1i}\xi_{2l}\Big)\Big(\xi_{1m}^*\xi_{2j}^*+
\xi_{1j}^*\xi_{2m}^*\Big)
\right]\;.
\eea
The charge fluctuations, then, reads
\be
\label{sigma_w0}
\sigma_L=n|\xi_{1L}|^2
|\xi_{1R}|^2+k|\xi_{2L}|^2|\xi_{2R}|^2+2nk|\xi_{1L}|^2|\xi_{2L}|^2\,,
\ee
and the decay rate $\Gamma^h_\gamma$, given by Eq.~(\ref{decayh}), is simply
\bea
\label{Gamma_w0}
\Gamma^h_\gamma=n|\xi_{1L}|^4+k|\xi_{2L}|^4+nk\left(|\xi_{1R}|^2|\xi_{2L}|^2+|\xi_{2R}|^2|\xi_{1L}|^2\right)\,,
\eea
where, of course $|\xi_{\alpha L}|^2+|\xi_{\alpha R}|^2=1$ is understood. So far we have used only $\omega=0$, which is a necessary but not sufficient condition 
for having spatially separated Fock state, as we will see in Sec.VIII.A.4.
Now, choosing the Fock state as follows 
\be
|\psi_F\rangle= \left(\xi_{1L} a^\dagger_L\right)^n\left(\xi_{2R} a^\dagger_R\right)^k|0\rangle\,,
\ee 
we are forced to put $\xi_{1i}=\delta_{iL}$ and $\xi_{2i}=\delta_{iR}$ in 
Eqs.~(\ref{PerDFock}, \ref{PerIFock}). This implies that the occupation numbers are $\langle n_L\rangle=n$ and $\langle n_R\rangle=k$, 
and the energy, the charge fluctuations, the visibility and the decay rate 
are simply given by
\bea
&&E=\frac{U}{2}\big(n(n-1)+k(k-1)\big)+\mu(n-k)\,,\\
&&\sigma_L=\sigma_R=0\,,\\
&&\alpha=0\,,\\
&&\Gamma_\gamma=|c_h|^2\,n(k+1)\,,
\eea
and, analogously, $\Gamma_\delta=|c_h|^2k(n+1)$ and 
$\Gamma_\beta=|c_h|^2(n+1)(k+1)$. The decay rate, in the Fock limit, is, 
therefore, $\propto N^2/4$, for $n\simeq k\simeq N/2$, 
and is due only to the coupling of the bath with
the hopping-like term.
\section{Coherent limit, $\omega=1$}
\label{sec.coherent}
For $\omega=1$ we have that
\bea
&&\phantom{.}_2F_1\left(-n,-k,1;1\right)=\left(
\ba{c}
n+k\\
n
\ea
\right)\,,\\
&&\sum_{\ell=0}^{k-1} \phantom{.}_2F_1\left(1-n,-\ell,1;1\right)=
\left(
\ba{c}
n+k-1\\
n
\ea
\right)\,,\\
%
&&\sum_{m=0}^{n-2}(1+m)\sum_{\ell=0}^{k-2} \phantom{.}_2F_1\left(-m,-\ell,1;1\right)=
\frac{n(n-1)}{k}\left(
\ba{c}
n+k-2\\
n
\ea
\right)\,.
\eea
We get, therefore, the following simple expressions
\bea
\label{coherent1}
&&\textrm{Per}(\Omega^{n,k})=(n+k)!\;,\\
&&\textrm{Per}({\cal O}^{n,k}_{1u})=\textrm{Per}({\cal O}^{n,k}_{1d})=(n+k-1)!
\;,\\
&&\textrm{Per}({\cal O}^{n,k}_{2u})=\textrm{Per}({\cal O}^{n,k}_{2d})=(n+k-2)!
\;.
\label{coherent3}
\eea
%
The coherent state is obtained when 
\bea
\xi_L\equiv \xi_{1L}=\xi_{2L}\,,\\
\xi_R\equiv \xi_{1R}=\xi_{2R}\,,
\eea
so that Eq.~(\ref{wf}) reduces to
\be
\label{wf.coh}
|\psi_C\rangle = \left(\xi_{L}a^\dagger_L+\xi_R a^\dagger_R\right)^{N}|0\rangle\,,
\ee
where $N=(n+k)$, the total number of bosons, and $|\xi_L|^2+|\xi_R|^2=1$. In the coherent case we obtain, therefore, simply
\bea
\label{PerDco} 
\textrm{Per}(D^{n,k}_{ij})&=&N!\left\{\delta_{ij}+N\xi_{i} \xi_{j}^*\right\}
\;,\\ 
\label{PerIco}   
\nonumber \textrm{Per}(I^{n,k}_{lijm})&=& N!\left\{\delta_{ij}\delta_{ml}+\delta_{jl}
\delta_{im}+ N \left(\delta_{lj}\xi_{i}\xi^*_{m}+\delta_{lm}\xi_{i}\xi^*_{j}+
\delta_{ij}\xi_{l}\xi^*_{m}+\delta_{im}\xi_{l}\xi^*_{j}\right) \right.\\
&&\left.+(N^2-N)\xi_{l}\xi_{i}\xi^*_{j}\xi^*_{m}\right\}\;.
\eea
Notice that Eqs.~(\ref{PerDco}) and (\ref{PerIco}) are valid also for a 
lattice, namely, in the presence of many sites, when the state is given by 
$|\psi\rangle =\left(\sum_i^{N_{s}}\xi_i a^\dagger_i\right)^N|0\rangle$. From Eq.~(\ref{PerDco}) we get that the occupation numbers are the following
\bea
\langle n_L\rangle=N|\xi_L|^2\,,\\
\langle n_R\rangle=N|\xi_R|^2\,.
\eea
Energy, charge fluctuations, visibility and decay rate are given by
\bea
\label{en.coh}
&&E=-\frac{t}{2} \left(\xi_L\xi_R^*+\xi_R\xi_L^*\right)N+\mu 
N(|\xi_L|^2-|\xi_R|^2)
+\frac{U}{2}\left(|\xi_L|^4+|\xi_R|^4\right)N(N-1)\,,\\
\label{sigmaC}
&&\sigma=\sigma_L=\sigma_R=N |\xi_L|^2|\xi_R|^2\,,\\
\label{alphaC}
&&\alpha=2|\xi_R\,\xi_L^*|\,,\\
\label{gammaC}
&&\Gamma_\gamma=N\left\{|c_L-c_R|^2\,|\xi_L|^2|\xi_R|^2+|c_h|^2\,|\xi_R|^4+
2\textrm{Re}[(c_L|\xi_L|^2-c_R|\xi_R|^2)c_h^*\xi_R\xi_L^*]\right\}\,,
\eea
and analogously, $\Gamma_\delta=N\left\{|c_L-c_R|^2\,|\xi_L|^2|\xi_R|^2+|c_h|^2\,|\xi_L|^4-2\textrm{Re}[(c_L|\xi_L|^2-c_R|\xi_R|^2)c_h^*\xi_R\xi_L^*]
\right\}$. 
What we found is, therefore, that, in the coherent regime, both 
the number fluctuations and the decay rate scale linearly with the number of bosons, no matter how the bath is coupled with the bosons. 
The couplings affect only the prefactor of the decay rate.
Before we conclude this section 
we present here 
a couple of interesting effects which are supposed to occur in the coherent-like regime.
\subsection{Self-trapping}
Let us consider small interaction so that the coherent state still well 
approximates the ground state or suppose that we can prepare the 
system in such a state. 
After defining the variables $z$, the relative charge imbalance, and $\phi$,
the phase difference between the left and right amplitudes,
\bea
&&z\equiv\left(\langle n_L\rangle- \langle n_R\rangle\right)/N=|\xi_L|^2-
|\xi_R|^2\,,\\
&&e^{i\phi}\equiv\frac{\xi_L\xi^*_R}{|\xi_L||\xi_R|}\,,
\eea
we can rewrite the total energy Eq.~(\ref{en.coh}), as follows
\be
\label{E_z_phi}
E=-\frac{t}{2} N\sqrt{1-z^2}\,\cos \phi+\frac{U}{4}N(N-1)(1+z^2)+\mu N z\,.
\ee
Eq.~(\ref{E_z_phi}) is 
the Josephson energy, where the hopping term is proportional to $\cos \phi$.
The effective Hamiltonian describing the evolution of $z$ is
${\cal H}=E/N$ which gives the following equations of motion \cite{augusto, augusto2,gurarie}
\bea
\dot{z}&=&\frac{t}{2}\sqrt{1-z^2}\sin\phi\,,\\
\dot{\phi}&=&\frac{Uz}{2}(N-1)-\frac{t z\cos\phi}{2\sqrt{1-z^2}}-\mu \,.
\eea
The Josephson current is, therefore, $I_J= N\dot{z}=\frac{tN^2}{2}         
\sqrt{1-z^2}\sin\phi$.
For $\mu \rightarrow 0$, we get the following non-trivial
fixed points
\bea
\label{selftrap}
z&=&\pm\frac{\sqrt{U^2(N-1)^2-t^2}}{U(N-1)}\,,\\
\phi&=&0,\pm \pi\,,
\eea
which means that, for a particular imbalance, the particles do not arrange 
themselves to reach the symmetric configuration.

\subsection{Attractive interaction: Symmetry breaking}
\label{sec.attractive}
We now consider the attractive interacting case, $U<0$, supposing that the 
state remains coherent also for negative $U$, not 
only strictly for zero interaction.
The total energy of a coherent state is given by Eq.~(\ref{en.coh}). 
The energy profile, for $\mu=0$ and $\phi=0$, has only one minimum at
\be
|\xi_L|^2=\frac{1}{2}, \;\; \textrm{for} \;\; \frac{t}{1-N}\le U\le 0\,,
\ee
while develops two minima at
\be
|\xi_L|^2=\frac{1}{2}\pm \frac{\sqrt{U^2(N-1)^2-t^2}}{2U(N-1)},\;\; 
\textrm{for} \;\; U<\frac{t}{1-N}\,.
\ee
For $U<\frac{t}{1-N}$, then, the ground state becomes twice degenerate 
and the charge imbalance, $z=2|\xi_L|^2-1$, is finite, given again by Eq.~(\ref{selftrap}).

\section{Large $N$ limit, 
$\omega\in (0,1]$}
In this section we derive the asymptotic behavior of the charge fluctuations 
and of the decay rate in the large $N$ limit. The result is valid as long as 
\be
|\omega|\gg N^{-1} 
\ee
and for $k=n=N/2$. In this case we obtain the following asymptotic 
behaviors for the permanents appearing in the correlators, Eqs.~(\ref{PerD}, \ref{PerI}),
\bea
\label{PerLim1}
&&{\textrm{Per}(\Omega^{n-1,n})} = {\textrm{Per}(\Omega^{n,n})}
\left[\frac{1-|\omega|+4n}{4(1+|\omega|)n^2}+O(n^{-3})\right]\,,\\
\label{PerLim2}
&&{\textrm{Per}({\cal O}^{n,n}_{1u})} = {\textrm{Per}(\Omega^{n,n})}
\left[\frac{|\omega|-1+4|\omega|n}{4\omega^*(1+|\omega|)n^2}+O(n^{-3})\right]
\,,\\
&&{\textrm{Per}({\cal O}^{n,n}_{2u})} = {\textrm{Per}(\Omega^{n,n})} 
\left[\frac{2\omega^2(|\omega|-1+|\omega|n)}{|\omega|^3(1+|\omega|)^2(2n-1)n^2}+O(n^{-4})\right]\,,\\
&&{\textrm{Per}(\Omega^{n-2,n})} = {\textrm{Per}(\Omega^{n,n})} 
\left[\frac{2-2|\omega|+2n}{(1+|\omega|)^2(2n-1)n^2}+O(n^{-4})\right]\,,\\
&&{\textrm{Per}(\Omega^{n-1,n-1})} = {\textrm{Per}(\Omega^{n,n})} 
\left[\frac{2}{(1+|\omega|)^2(2n-1)n}+O(n^{-4})\right]\,,
\label{PerLim5}
\eea 
and
$\textrm{Per}({\cal O}^{n,n}_{1d})=\textrm{Per}({\cal O}^{n,n}_{1u})^*$,  
$\textrm{Per}(\Omega^{n,n-1}) = \textrm{Per}(\Omega^{n-1,n})$, 
$\textrm{Per}(\Omega^{n,n-2}) = \textrm{Per}(\Omega^{n-2,n})$.
Notice that for $\omega=1$ only the leading terms survive and we recover the 
exact results for the coherent states, Eqs.~(\ref{coherent1}-\ref{coherent3}). 

Using these relations we can calculate 
the charge fluctuation and the decay rate.
For simplicity we now choose all $\xi$'s real so that $\omega$ is also real. 
Since we are considering the simple case with $k=n$, namely 
a symmetrically occupied double well, we can use the following parametrization 
\bea 
\label{param1}
\xi_{1L}=\xi_{2R}=\frac{1}{\sqrt{2}} \sqrt{1+\sqrt{1-\omega^2}}\,,\\ 
\xi_{2L}=\xi_{1R}=\frac{1}{\sqrt{2}} \sqrt{1-\sqrt{1-\omega^2}}\,,   
\label{param2}
\eea                                                                           
which fulfils the normalization conditions and the definition of the single 
particle overlap. We have also checked that, for $\mu=0$ and repulsive 
interaction, $U\ge 0$, the balance conditions ($k=n$ and $\xi_{1L}=\xi_{2R}$, 
$\xi_{1R}=\xi_{2L}$) naturally minimize the total energy.   
After this parametrization we can calculate, for instance, $\sigma$ and 
$\Gamma_\gamma$, and, after expanding in $N^{-1}$, we obtain, at the leading 
order, 
\bea
\label{sigmaL-}
&&\sigma_L = \sigma_R=\sigma\simeq \frac{\omega N}{4} \,,\\ 
&&\Gamma_{\gamma}\simeq\frac{N}{4}\left|\frac{c_h}{\sqrt{\omega}}+\sqrt{\omega}(c_L-c_R)\right|^2\,,
\eea
and, analogously, 
$\Gamma_{\delta}\simeq\frac{N}{4}\left|\frac{c_h}{\sqrt{\omega}}-
\sqrt{\omega}(c_L-c_R)\right|^2$. 
Calculating the total energy, Eq.~(\ref{totalE}), with the asymptotics written 
above, Eqs.~(\ref{PerLim1})-(\ref{PerLim5}), we obtain 
\bea
\label{toten}
E&=&\nonumber\frac{U
\Big(8+4\omega(\omega^2-3)-2N+
2\omega(4+\omega-2\omega^2)N+N^2(1+\omega)^2(N+\omega-3)\Big)}
{4(1+\omega)^2(N-1)}+\\
&&+\frac{t}{4\omega}
\Big(1+\omega^2-2\omega(N+1)\Big)\,,
\eea
and imposing $\delta_\omega E=0$, we get the minimum energy when 
\be
\label{wmin}
\omega_o\simeq \frac{\sqrt{t}}{\sqrt{t+N U}}\,,
\ee
which is consistent with the limit $\omega\sim N^{-1/2}\gg N^{-1}$ 
the asymptotic is based on. 
As a result, the decay rate at $U=0$, is
\be
\Gamma_{\gamma}\simeq
\frac{N}{4}\left|c_L-c_R+c_h\right|^2
\ee
and $\Gamma_{\delta}\simeq\frac{N}{4}\left|c_L-c_R-c_h\right|^2$. 
For $U/t$ of order one and for large $N$, 
instead, the leading term is given by 
\be
\Gamma_{\gamma} \simeq \frac{N\,|c_h|^2}{4\,\omega_o}\simeq \frac{|c_h|^2}{4}\sqrt{\frac{{U}}{{t}}}\,N^{3/2}\,.
\ee 
We know that $\Gamma_{\gamma} \simeq N^2|c_h|^2/4$ for the Fock states,
meaning that when $\omega_o\sim N^{-1}$, which implies $U \sim t N$,
namely, when the interaction is of order $N$, 
we enter the Fock regime.\\
The charge fluctuations, far from the Fock regime, are
\be
\sigma\simeq \frac{\omega_o N}{4} \simeq \frac{1}{4}\sqrt{\frac{t}{U}}\,N^{1/2}\,.
\ee
For $U=0$, namely $\omega_o=1$, we recover the result of Sec.~\ref{sec.coherent}, with $|\xi_{L}|=|\xi_R|=1/\sqrt{2}$, since, without interaction, the 
coherent state is the ground state. 
Finally, making an expansion in $N^{-1}$ of the total energy, 
Eq.~(\ref{toten}), calculated at the minimum, Eq.~(\ref{wmin}), we get 
\be
\label{enmin}
E_o=-\frac{t}{2}N+\frac{1}{4}UN(N-2)+\frac{1}{2}\sqrt{tUN}+O(1)\,,
\ee
which, in spite of its simplicity, is a very good approximation of the 
ground state energy for intermediate interaction, 
i.e. $\frac{1}{N}\ll \frac{U}{t}\ll N$.
By exactly diagonalizing the Hamiltonian for different values of $N$, 
in fact, one can easily check that 
Eq.~(\ref{enmin}) deviates from the exact ground level $E_{ex}$ by  
an error, 
${E_o-E_{ex}}$, of order 
$O(1)\approx \left(\frac{t}{2}+\frac{U}{16}\right)$ 
(relative error of order $O(N^{-2})$).
Notice that the coherent energy, $E_C=-\frac{t}{2}N+\frac{U}{4}N(N-1)$, 
deviates from the exact one by an error $O(N)\approx \frac{U}{4}N$ 
and the Fock energy, $E_F=\frac{U}{4}N(N-2)$, by an error 
$O(N)\approx \frac{t}{2}N$ (both relative errors of order 
$O(N^{-1})$).
As a result, Eq.~(\ref{enmin}) can be considered as a non-perturbative
analytical expression of the ground state energy for intermediate interaction.

Let us finally, consider the visibility, Eq.~(\ref{vis2}). This quantity 
is equal to $1$ in the 
coherent-like state ($U= 0$) and decreases by increasing the repulsive 
interaction ($U>0$). 
Using Eq. (\ref{PerD}) and Eqs.~(\ref{PerLim1}),~(\ref{PerLim2}), 
for $n=k=N/2$, and Eqs.~(\ref{param1}),~(\ref{param2}), we obtain, in fact, 
for $0\le U/t\ll N$, at the leading orders,
\be
\label{alpha}
\alpha \simeq 1-\frac{(\omega_o-1)^2}{2\omega_o N} \,\simeq\, 1-
\frac{\left(\sqrt{t}-\sqrt{t+NU}\right)^2}{2N\sqrt{t(t+NU)}}.
\ee
This result for the coherence visibility is in a very good agreement with 
numerical results \cite{mazzarella}. 
To conclude, both the energy, Eq.~(\ref{enmin}), and the visibility, 
Eq.~(\ref{alpha}), have been successfully compared with exact diagonalization 
results, 
validating therefore our ansatz of the interpolating 
state in the regime of repulsive interaction. 

\section{Entanglement for $U\ge 0$}
In this section, we conclude our study calculating the entanglement entropy 
and the Fisher information of our permanent state.  
\subsection{Entropy}
Given the wavefunction (\ref{wf}), we can derive the reduced density 
matrix for the left site, for instance, by tracing out the right one 
(for a review, see Ref.~\cite{amico} and references therein)
\be
\hat\rho=\frac{1}{\langle \psi|\psi\rangle}\,\textrm{Tr}_R\left(|\psi\rangle \langle \psi|\right), 
\ee
where $\textrm{Tr}_R$ is the trace over the right site. More explicitly, we have to perform the following summation
\be
\hat\rho=\frac{1}{\langle \psi|\psi\rangle} \sum_{m=0}^{n+k} \frac{1}{m!}\langle 0| (a_R)^m |\psi\rangle \langle \psi| (a_R^\dagger)^m |0 \rangle\,.
\ee
This is a diagonal $(n+k+1)\times (n+k+1) $ matrix, $\rho_{\ell\ell^\prime}=\rho_\ell\delta_{\ell\ell'}$, 
whose diagonal elements can be written analytically as follows
\be
\label{rho1}
\rho_\ell=\frac{|\xi_{2L}|^{2n}|\xi_{1R}|^{2\ell}|\xi_{2R}|^{2(k-\ell)}\,(n+k-\ell)! \,k!\,
\left|_2F_1\left(-n,-\ell,1-\ell+k;\frac{\xi_{1L}\xi_{2R}}{\xi_{1R}\xi_{2L}}\right)\right|^2}{n!\,\ell!\,\left((k-\ell)!\right)^2\, 
_2F_1\big(-n,-k,1;|\xi_{2L}\xi_{1L}^*+\xi_{2R}\xi_{1R}^*|\big)}\,,
\ee
for $0\le \ell< k$ and
\be
\label{rho2}
\rho_\ell=\frac{|\xi_{1L}|^{2(\ell-k)}|\xi_{2L}|^{2(n+k-\ell)}|\xi_{1R}|^{2k}\,\ell! \,n!\,
\left|_2F_1\left(\ell-n-k,-k,1+\ell-k;\frac{\xi_{1L}\xi_{2R}}{\xi_{1R}\xi_{2L}}\right)\right|^2
}{k!\,(n+k-\ell)!\,\left((\ell-k)!\right)^2 \,
_2F_1\big(-n,-k,1;|\xi_{2L}\xi_{1L}^*+\xi_{2R}\xi_{1R}^*|\big)}\,, 
\ee
for $k\le \ell \le n+k$. The $\xi$'s are related by normalization conditions, Eq.~(\ref{normaliz}). We can therefore calculate the von Neumann entropy
\be
\label{entropy}
S=-\sum_{\ell=0}^{n+k}\rho_\ell\log_2 \rho_\ell 
\ee
with $\log_2$ the logarithm to base $2$. We have shown that also the reduced density matrix and the entropy can be written in terms of Jacobi polynomials of $\omega$ and of the ratio $\frac{\xi_{1L}\xi_{2R}}{\xi_{1R}\xi_{2L}}$.
\subsubsection{Fock limit}
In the Fock limit, $\omega=0$, choosing $\xi_{1i}=\delta_{iR}$ and 
$\xi_{2i}=\delta_{iL}$ (the other choice is equivalent but one should take 
care of the ratio $\frac{\xi_{1L}\xi_{2R}}{\xi_{1R}\xi_{2L}}$, the argument of the Jacobi polynomial appearing in the numerator of the density matrix), 
from Eqs.~(\ref{rho1}), (\ref{rho2}), we get
\be
\rho_\ell=\delta_{\ell k} \,.
\ee
As a consequence the entropy is simply $S=0$.
\subsubsection{Coherent limit}
If $\xi_{1L}=\xi_{2L}$ and $\xi_{1R}=\xi_{2R}$, then $\omega=1$, and 
the reduced density matrix becomes simply
\be
\label{rho_co}
\rho_\ell=\left(\ba{c} N\\\ell\ea \right)|\xi_L|^{2\ell}|\xi_R|^{2(N-\ell)}\,,
\ee
where, of course, $N=n+k$ and $|\xi_R|^2=(1-|\xi_L|^2)$.
In this case we can calculate the asymptotic behavior of the von Neumann 
entropy for $N\gg 1$, since the binomial distribution, Eq.~(\ref{rho_co}), 
approaches the gaussian one
\be
\rho_\ell\simeq \frac{1}{\sqrt{2\pi N|\xi_L|^2|\xi_R|^2}}
\,\exp\left[-\frac{(\ell-N|\xi_L|^2)^2}{2N|\xi_L|^2
|\xi_R|^2}\right]\,,
\ee
obtaining the following asymptotic behavior for the entropy \cite{jstat} 
\be
\label{Sco}
S\simeq \frac{1}{2}
\log_2\left(2\pi e N|\xi_L|^2|\xi_R|^2\right)\,.
\ee

\subsubsection{Intermediate case, with $n=k$}
\begin{figure}[h!]
\centering
\includegraphics[width=7cm]{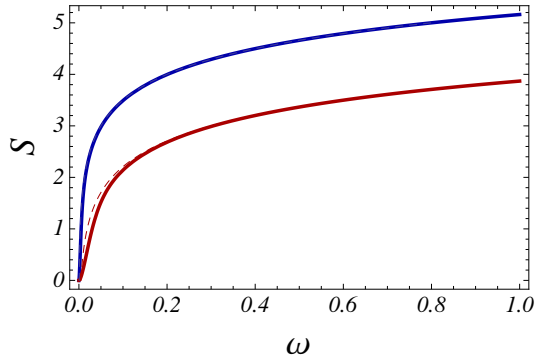}
\caption{(Color online) Entanglement entropy, $S$, in the repulsive regime, 
as a function of $\omega$,
for $N=300$ (upper solid line) and $N=50$ (lower solid line).
The dashed lines are the approximated behaviors given by Eq.~(\ref{Sapom}),
for $N=300$ (upper dashed line, which cannot be distinguished from 
the solid line) 
and $N=50$ (lower dashed line), 
which deviates from the solid line only for small $\omega$.}
\label{fig.Som}
\end{figure}

Now let us consider the case of $k=n=N/2$ and Eqs.~(\ref{param1}),~(\ref{param2}). Applying Eqs.~(\ref{rho1}), (\ref{rho2}) and (\ref{entropy}) we get an entropy which is well approximated, for $\omega\gg 1/N$, by
\be
\label{Sapom}
S\approx\frac{1}{2}\log_2\left(\pi e \,\omega \frac{N}{2}\right)\,,
\ee
as one can see from Fig.~(\ref{fig.Som}). At $\omega=\omega_o$, for $U/t$ of order one, we get $S\sim \frac{1}{4}\log_2(N)+$ const. 
 
\subsubsection{A special case: an almost maximally entangled state}
\begin{figure}[h!]
\centering
\includegraphics[width=7cm]{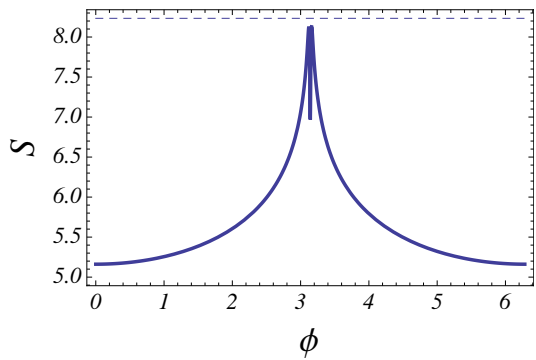}
\caption{(Color online) Entanglement entropy for the state in Eq.~(\ref{2cond}), with 
$n=k=N/2$, as a function of $\phi$. For $N=300$, the entropy reaches 
its maximum at $\phi=\pi\pm 0.02$ which is $S(\phi)\approx 8.13$. 
The dashed line is the upper limit entropy, $\log_2(N+1)$, which, 
for $N=300$, is $\approx 8.23$.
Exactly at $\phi=\pi$ there is a narrow dip where $S(\phi=\pi)\approx 6.96$, 
nicely approximated by Eq.~(\ref{S2cond}), which gives $\approx 7.23$.  
At $\phi=0$ and $2\pi$, the entropy coincides with Eq.~(\ref{Sco}), giving  
$S(\phi=0)\approx 5.16$.}
\label{fig.Sphi}
\end{figure}

In this paragraph we will consider a simple 
state which has the same functional form of Eq.~(\ref{wf}), 
with same amplitudes in modulus, 
obtained by a phase deformation of the balanced coherent-like state. 
We will show that the phase can be tuned in order to 
get an almost maximally entangled state. 
Let us consider, therefore, Eq.~(\ref{wf}), with
\be
\xi_{1L}=\xi_{2L}=
\xi_{1R}=e^{-i\phi}\xi_{2R}=1/\sqrt{2}\,,
\ee 
such that the not normalized state reads
\be
\label{2cond}
|\psi\rangle=\frac{1}{\sqrt{2^{n+k}}}\left(a_L^\dagger+a_R^\dagger\right)^{n}\left(a_L^\dagger+e^{i\phi}a_R^\dagger\right)^{k}|0\rangle\,,
\ee
then the overlap parameter is given by
\be
\omega=
\frac{1}{2}(1+e^{i\phi})\,.
\ee
When $\phi=0$ we recover the coherent state. Using Eqs.~(\ref{rho1}),
~(\ref{rho2}) and (\ref{entropy}) one can verify that the state in 
Eq.~(\ref{2cond}) is more entangled than the fully delocalized coherent state,
i.e. 
$S(\phi\neq 0)>S(\phi=0)$, as we can see in Fig.~\ref{fig.Sphi}.\\

In particular, for $\phi=\pi$ the single particle overlap is $\omega=0$, like the Fock state. Defining 
\bea
a_1^\dagger=(a_L^\dagger+a_R^\dagger)/\sqrt{2}\,,\\
a_2^\dagger=(a_L^\dagger-a_R^\dagger)/\sqrt{2}\,,
\eea
the state in Eq.~(\ref{2cond}) is, indeed, a Fock state in the new 
representation, $|\psi\rangle=(a_1^\dagger)^{n}
(a_2^\dagger)^{k}|0\rangle$, 
very unstable under coupling to an external bath,
in fact $\Gamma^h_\gamma\simeq nk/2$ (see Eq.~(\ref{Gamma_w0})), 
and, at the same time, with a quite large number fluctuations, 
$\sigma\simeq nk/2$, (see Eq.~(\ref{sigma_w0})). 
For $\phi=\pi$ and $k=n=N/2$, the reduced density matrix is simply given by 
\bea
\label{rho3}
&&\rho_{2\ell}=\frac{(2\ell)!(2n-2\ell)!}{4^n\left(\ell!\right)^2(\left(n-\ell\right)!)^2}\,,\\
&&\rho_{2\ell+1}=0\,.
\eea
Studying the profile of the reduced density matrix, Eq.~(\ref{rho3}), one can 
check that, it can be approximated as 
$\rho_{2\ell}\approx 1/n=2/N$, yielding, therefore, the following 
approximated value for the entropy
\be
\label{S2cond}
S\approx \log_2(N)-1\,.
\ee
The state Eq.~(\ref{2cond}) with $\phi=\pi$ and $n=k$, is, therefore, almost 
double entangled with respect to the coherent state ($\phi=0$), as one can 
see by comparing Eq.~(\ref{S2cond}) with 
Eq.~(\ref{Sco}). In conclusion, the state in Eq.~(\ref{2cond}), 
at $\phi\approx \pi$, is almost maximally entangled since the entropy 
approaches the upper limit for $N$ bosons, i.e. $\log_2(N+1)$, 
as shown in Fig.~\ref{fig.Sphi}. What shown is a simple example of how the 
entanglement can increase by losing coherence.

\subsection{Fisher information}
The Fisher information is defined by \cite{wootters,pezze,weiss,mazzarella}
\be
\label{F}
F=\frac{\langle(n_L-n_R)^2\rangle-\langle(n_L-n_R)\rangle^2}{N^2}\,,
\ee
which can be written as follows
\be
\label{F2}
F=\frac{1}{N^2}\left\{\sigma_L+\sigma_R-2\big(\langle n_Ln_R\rangle-\langle n_L\rangle\langle n_R\rangle\big)\right\}\,.
\ee
If we consider the case with $n=k=N/2$, after choosing $\xi$ real and parametrizing them as in Eqs.~(\ref{param1}), (\ref{param2}) we get
\be
F\simeq  \frac{\omega^2}{2}\,,
\ee
for $\omega\ll N^{-1}$, and
\be
\label{Fomega}
F\simeq \frac{\omega}{N}\,,
\ee
for $\omega\gg N^{-1}$. The overlap of the bosons can be seen, therefore, as 
the parameter which encodes the quantum information. In the latter case, 
i.e. $\omega\gg N^{-1}$, we find, in fact, that, at least at the leading term,
\bea
\label{nlnr}
\langle n_Ln_R\rangle-\langle n_L\rangle\langle n_R\rangle \simeq -\sigma
\eea
where $\sigma=\sigma_L=\sigma_R$. Equation (\ref{nlnr}) becomes 
exact in the coherent limit, $\omega=1$. 
This implies a very interesting relation, 
valid at least for $\omega\gg N^{-1}$,
\be
\label{Fsigma+}
F\simeq \frac{4\,\sigma}{N^2}\,.
\ee
For finite interaction and at the ground state energy, we know that $\omega_o\simeq \sqrt{t/(NU)}$, therefore, we get 
\be
F\simeq \sqrt{\frac{t}{U}}\,N^{-3/2}\,.
\ee
We notice that the Fisher information and the decay time $\tau_\gamma\equiv 1/\Gamma_\gamma$, for finite $U/t$, behave in the same way, namely they are proportional, i.e. $F\simeq \tau_\gamma {|c_h|^2}/{4}$, 
at least for a symmetrically occupied double well and large $N$.
We can say, therefore, that the greater is the Fisher information, 
the longer the state survives under coupling to an external environment.

\section{Attractive interaction: the Schr\"odinger cat}
\begin{figure}[h!]
\centering
\includegraphics[width=4.2cm]{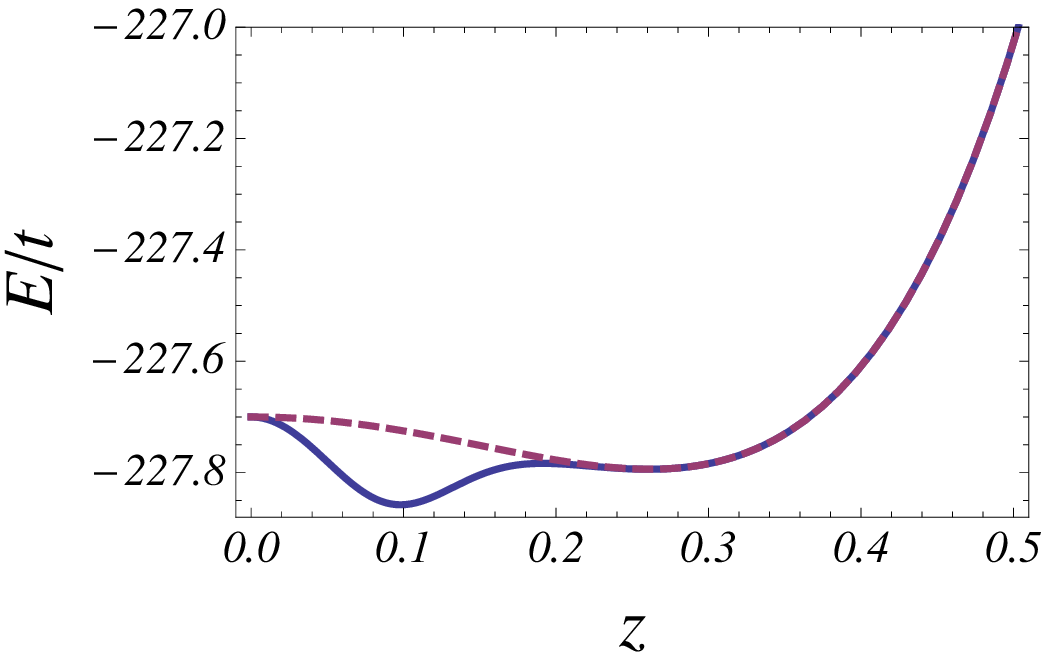}\hspace{-0.4cm}
\includegraphics[width=4.2cm]{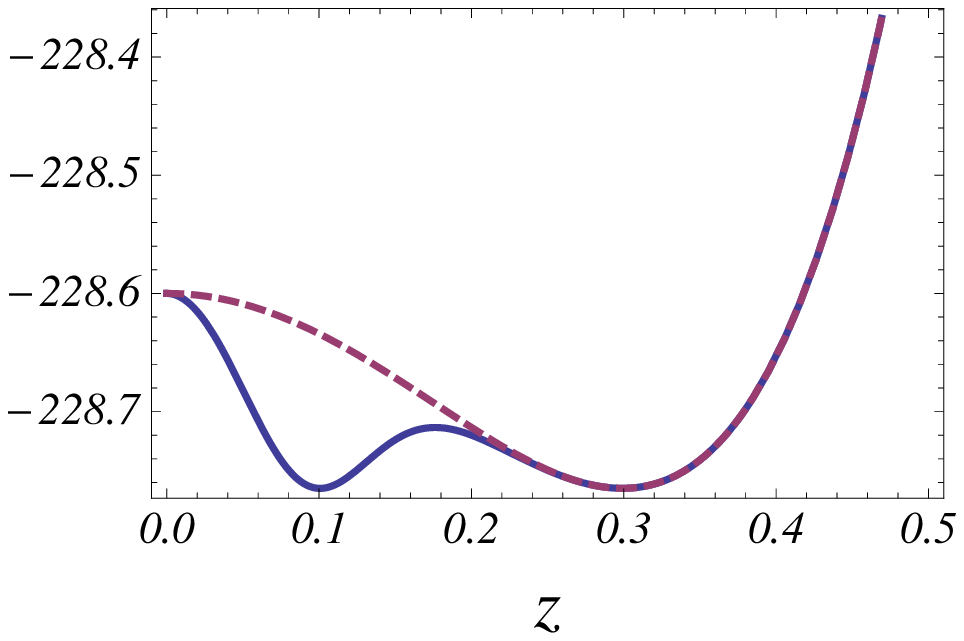}\hspace{-0.4cm}
\includegraphics[width=4.2cm]{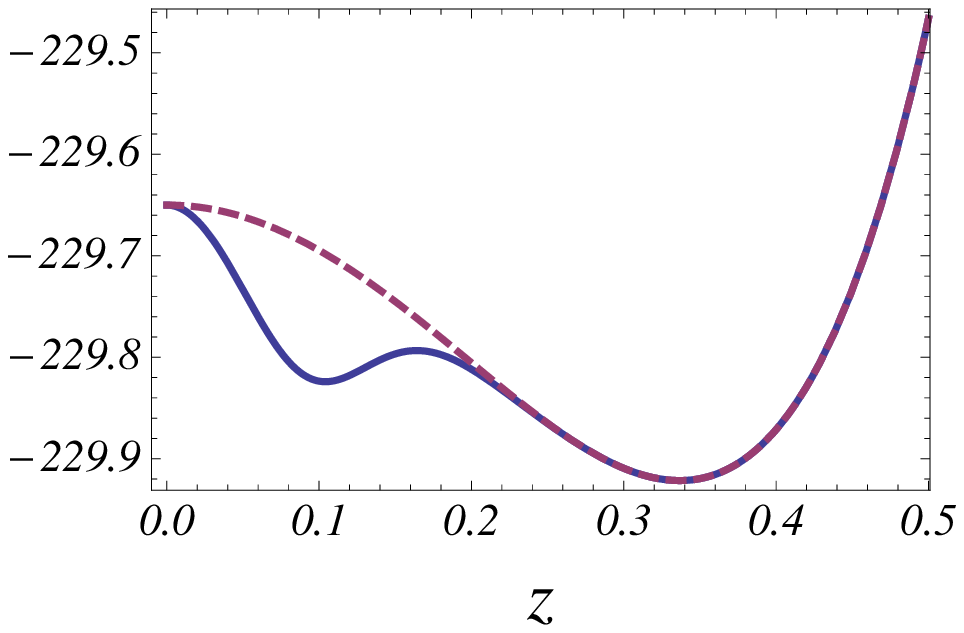}\hspace{0.05cm}
\includegraphics[width=4cm]{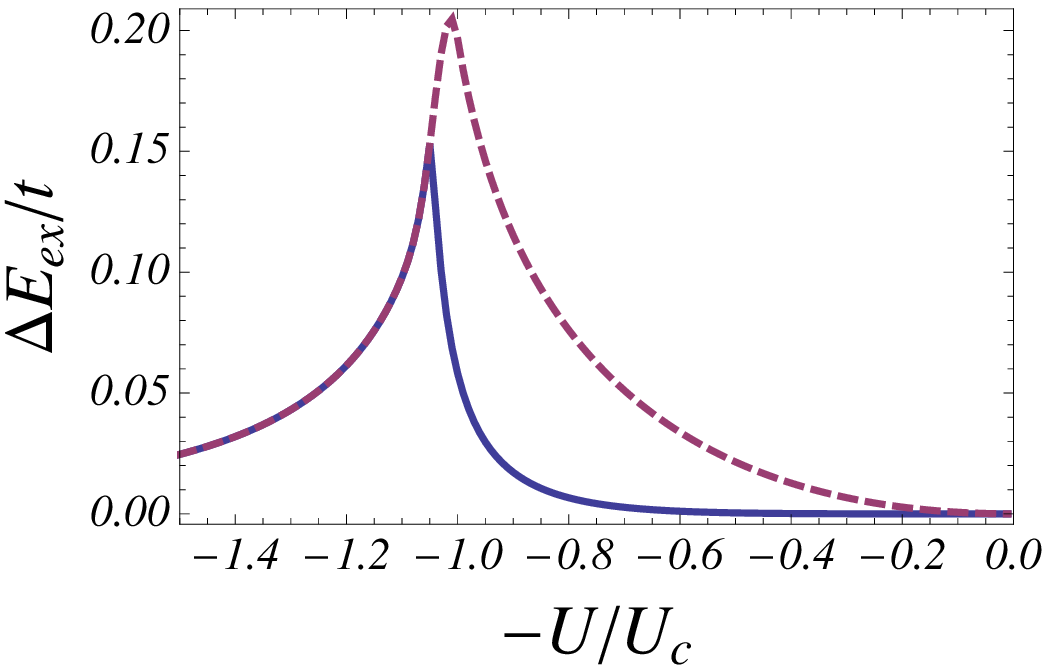}
\caption{(Color online) (First three plots) Energy $E$ as a function of $z$, Eq.~(\ref{enz1}), for $N=300$ and $U=x\,U_c$, with $x=1.036, 1.048, 1.062$ from left to right. There is a critical 
interaction $U_c^*<U_c$ at which the global minimum of $E$ jumps 
from one value of $z$ the another. The red dashed line is $E_s$ as a function 
of $z$, Eq.~(\ref{enz2}), for the same values of $N$ and $U$. 
The energies from exact diagonalization, for those values of interaction,  
are, respectively, $E_{ex}=-227.97, -228.92, -230.05$ in units of $t$. 
(Last plot) Energy differences, between the global minimum of $E$ and the exact ground state energy $E_{ex}$, i.e. $\Delta E_{ex}=\min(E)-E_{ex}$, 
(solid blue line) and between the global minimum of $E_s$ and $E_{ex}$, 
i.e. $\Delta E_{ex}=\min(E_s)-E_{ex}$, (red dashed line), for $N=300$. 
We observe  
that $\min(E)$ is much better than $\min(E_s)$ for $U>U_c^*$, and 
deviates from exact ground level mainly for $U\simeq U_c^*$, although, 
the relative error at that point is still very small, less than $0.07\%$.}
\label{fig.E123}
\end{figure}

\begin{figure}[h!]
\centering
\includegraphics[width=5cm]{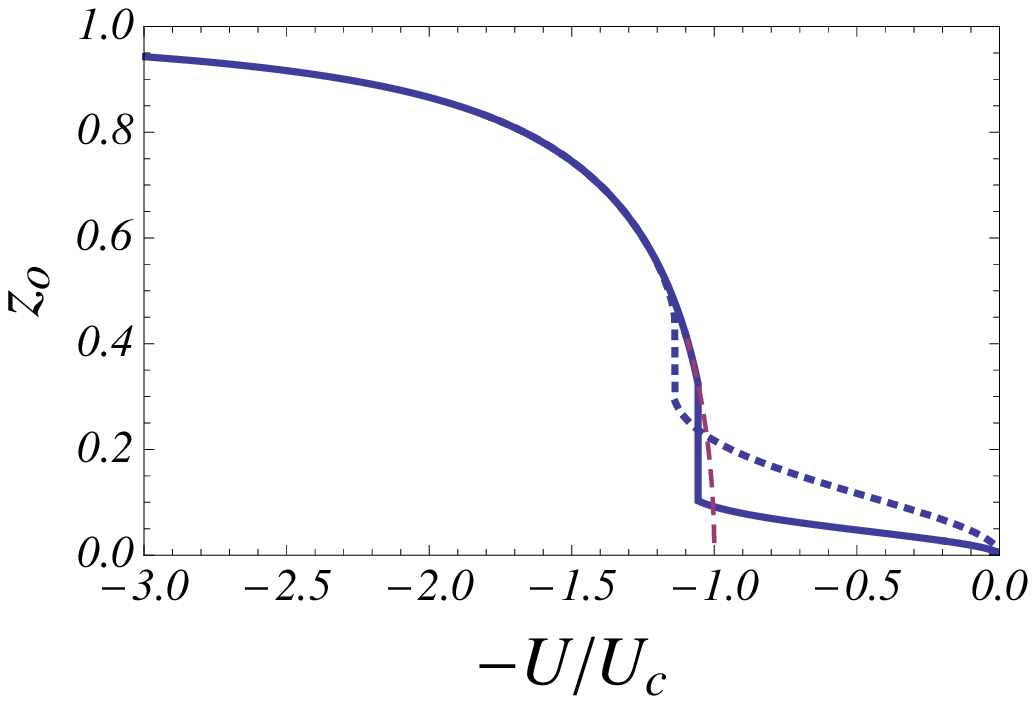}
\hspace{0.25cm}
\includegraphics[width=5cm]{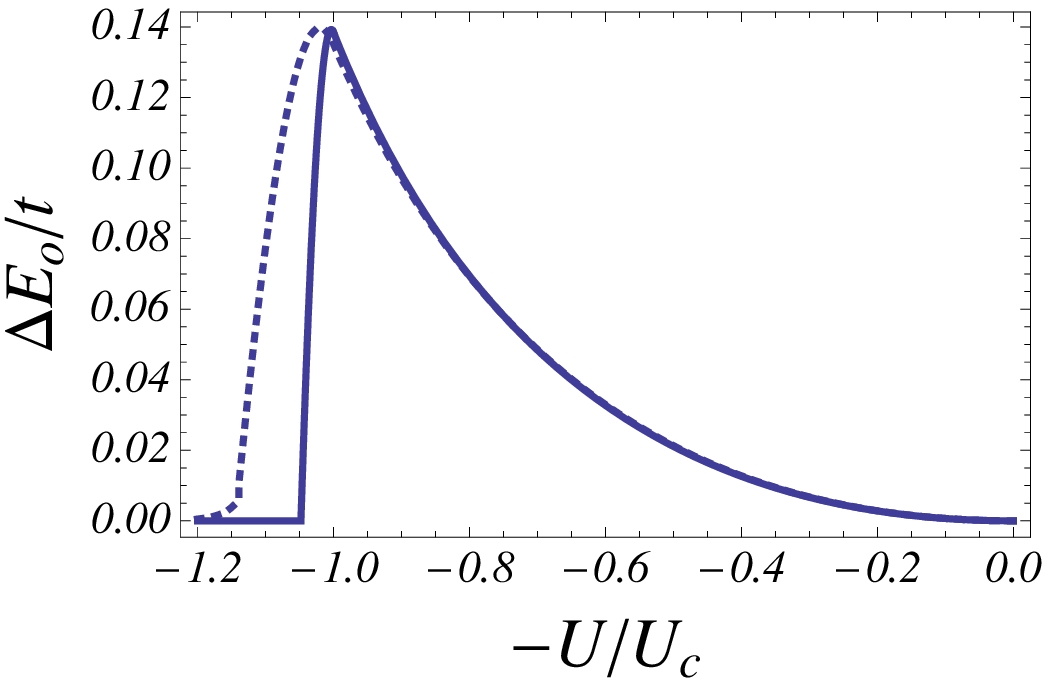}\hspace{0.3cm}
\includegraphics[width=5cm]{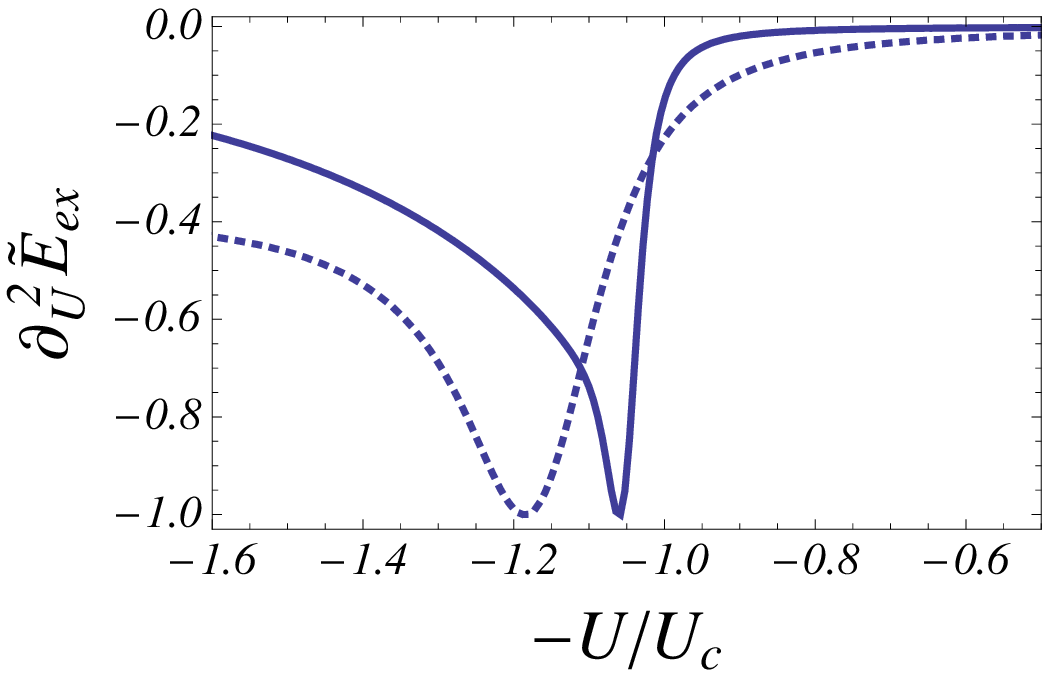}
\caption{(Color online) (Left plot) Imbalance $z_o$, calculated as the global 
minimum of $E$ given 
by Eq.~(\ref{enz1}), as a function of $U$ in units of $(-U_c)$, for $N=300$ 
(solid line) and $N=50$ (dotted line). 
The red long-dashed line is the plot of $z_o$ given by Eq.~(\ref{z}), 
i.e. the minimum of $E_s$ given by Eq.~(\ref{enz2}). We can see that at $U=U_c^*<U_c$ 
a discontinuity of $z_o$ occurs. 
(Central plot) Energy difference $\Delta E_o=\min(E_s)-\min(E)$, 
between the global minimum of $E_s$ and the global minimum of $E$, as a function of $U$, for $N=300$ (solid line) and $N=50$ (dotted line). 
$\Delta E_o$ is always non-negative, large for $0>U>U_c^*$ and almost zero 
for $U\le U_c^*$. 
(Right plot) Second derivative of the ground state energy 
$E_{ex}$, obtained by exact diagonalization with respect to $U$, i.e. 
${\partial^2 E_{ex}}/{\partial U^2}$, as a function of $U$, for $N=300$ 
(solid line) and $N=50$ (dotted line). The energies $E_{ex}$ have been 
rescaled for a better comparison: $\tilde{E}_{ex}=E_{ex}/(294.55\,t)$, for $N=300$, and 
$\tilde{E}_{ex}=E_{ex}/(5.68\,t)$, for $N=50$.}
\label{fig.zU}
\end{figure}

In this section we will study the evolution from a coherent-like state to a 
$NOON$ state, namely $|N\rangle_L|0\rangle_R+|0\rangle_L|N\rangle_R$, 
reached in the limit $U\rightarrow-\infty$. 
As we have seen before, this regime seems to be characterized by a 
symmetry breaking, as soon as $U<U_c$, where
\be 
\label{Uc}
U_c=-\frac{t}{(N-1)}\,.
\ee
We will consider, 
therefore, the following ``cat'' state, superposition of two unbalanced 
coherent-like states, 
\be
\label{cat}
|\psi_{@}\rangle=|\psi_L\rangle + |\psi_R\rangle \,,
\ee
with 
\bea
|\psi_L\rangle=\left(\xi^{>} a_L^\dagger+\xi^{<} a_R^\dagger\right)^N|0\rangle,\\
|\psi_R\rangle=\left(\xi^{<} a_L^\dagger+\xi^{>} a_R^\dagger\right)^N|0\rangle.
\eea
where $|\xi^{>}|\ge |\xi^{<}|$, and $|\xi^{>}|^2+|\xi^{<}|^2=1$, 
meaning that $|\psi_L\rangle$ is more weighted on 
the left site while $|\psi_R\rangle$ on the right site. 
After defining the charge imbalances for $|\psi_L\rangle$ and $|\psi_R\rangle$, one opposite to the other,
\be
z= z_L=-z_R=|\xi^>|^2-|\xi^<|^2\,,
\ee
the energy in the quantum coherent regime, can be written as in Eq.~(\ref{E_z_phi}) (with $\mu=0$). 
Imposing $\partial E/\partial z=0$ and $\partial E/\partial \phi=0$, we get 
two minima, for any $\phi=2n\pi$, 
\be
\label{z}
z_o=\pm \frac{\sqrt{U^2(N-1)^2-t^2}}{U(N-1)},
\ee 
as soon as $U<U_c$. The two solutions correspond to the state $|\psi_L\rangle$ or to the state $|\psi_R\rangle$. 
Actually, the energy is even reduced, as we will show in what follows, 
by taking the symmetric linear combination of those two states, which leads 
to the cat state written in Eq.~(\ref{cat}). 
It describes a superposition of a cat sitting on the 
left site but with his tail on the right one and viceversa. When the cat 
withdraws the tail we get the $NOON$ state. \\
In order to calculate the quantities of interest on the cat state, we first 
write the following correlators, with 
$|\psi_{s}\rangle=\{|\psi_{L}\rangle,|\psi_{R}\rangle\}$,
\bea
\label{perOmss'}
&&\langle \psi_{s}|\psi_{s'}\rangle=N!\,b_{s,s'}^N\equiv 
\textrm{Per}(\Omega^{N,0}(s,s')) \,,\\
\label{perDss'}
&&\langle \psi_{s}| a_i a_j^\dagger|\psi_{s'}\rangle=
N!\left(\delta_{ij}b_{s,s'}^N+Nb_{s,s'}^{N-1}\xi^{s}_i\xi^{s'*}_j\right)
\equiv\textrm{Per}(D^{N,0}_{ij}(s,s')) \,, \\
\label{perIss'}
&&\nonumber\langle \psi_{s}| a_l a_i a_j^\dagger a_m^\dagger|\psi_{s'}\rangle
= \delta_{lm}\textrm{Per}(D^{N,0}_{ij}(s,s'))+\delta_{lj}
\textrm{Per}(D^{N,0}_{im}(s,s'))+\delta_{im}\textrm{Per}(D^{N,0}_{lj}(s,s'))\\
&&\phantom{\langle \psi_{s}| a_l a_i a_j^\dagger a_m^\dagger|\psi_{s'}\rangle}
-\delta_{lj}\delta_{im}\textrm{Per}(\Omega^{N,0}(s,s'))+
N^2\xi^{s}_l\xi^{s'*}_m\textrm{Per}(D^{N-1,0}_{ij}(s,s')) \,,
\eea
where $s=L,R$ and
\bea
&&\xi^L_L=\xi^R_R=\xi^>\,,\\
&&\xi^L_R=\xi^R_L=\xi^<\,,\\
&&b_{L,L}=b_{R,R}=1\,,\\
&&b_{L,R}=b_{R,L}=\xi^> \xi^{<*}+\xi^< \xi^{>*}\,.
\eea 
The normalization of the cat state is, therefore, given by
\bea
\nonumber\langle\psi_{@}|\psi_{@}\rangle&=&2 N!\left[1+(\xi^>\xi^{<*}
+\xi^<\xi^{>*})^N
\right]=2 N!\left[1+\left(\sqrt{1-z^2}\cos\phi\right)^N\right]\Big{|}
_{\phi=2n\pi}\\
&=&2 N!\left[1+\left(\sqrt{1-z^2}\right)^N\right].
\eea
Without loss of generality, choosing $\xi^>$, $\xi^<$ real, 
from Eqs.~(\ref{perDss'}),~(\ref{perIss'}), we get the following quantities,  
useful to calculate energy, Fisher information, number fluctuations and decay 
rates, all in terms of $z$, the imbalance parameter,
\bea
\label{aLaR}
&&\hspace{-0.5cm}\langle a_L^\dagger  \,a_R \rangle_@ =\langle a_R^\dagger  \,a_L \rangle_@
\equiv 
\frac{\langle \psi_@ |a_R^\dagger a_L|\psi_@\rangle}
{\langle \psi_@|\psi_@\rangle}
= \frac{N}{2}\left(\frac{\sqrt{1-z^2}+\left(\sqrt{1-z^2}\right)^{N-1}}{1+\left(\sqrt{1-z^2}\right)^N}\right)\,,\\
&&\hspace{-0.5cm}\langle n_L\rangle_@ =\langle n_R\rangle_@ \equiv 
\frac{\langle \psi_@ |a_L^\dagger a_L
|\psi_@\rangle}{\langle \psi_@|\psi_@\rangle}
=\frac{N}{2}\,,\\
&&\hspace{-0.5cm}\langle n_L^2\rangle_@ =\langle n_R^2\rangle_@ \equiv
\frac{\langle \psi_@ |
a_L^\dagger a_La_L^\dagger a_L|\psi_@\rangle}{\langle \psi_@|\psi_@\rangle}
=\frac{N}{2}+\frac{N}{4}(N-1)\left(1+\frac{z^2}
{1+\left(\sqrt{1-z^2}\right)^N}\right)\,,\\
\label{nLnR}
&&\hspace{-0.5cm}\langle n_L n_R\rangle_@ \equiv 
\frac{\langle \psi_@ |
a_L^\dagger a_La_R^\dagger a_R|\psi_@\rangle}{\langle \psi_@|\psi_@\rangle}
=\frac{N}{4}(N-1)\left(1-\frac{z^2}{1+\left(\sqrt{1-z^2}\right)^N}\right)\,,\\
&&\hspace{-0.5cm}\langle n_L a_L^\dagger  \,a_R \rangle_@ =
\langle a_R^\dagger  \,a_L n_L \rangle_@=\langle n_R a_R^\dagger \,a_L 
\rangle_@ =\langle a_L^\dagger  \,a_R n_R \rangle_@=
\frac{(N+1)}{2}\langle a_L^\dagger  \,a_R \rangle_@\,,\\
&&\hspace{-0.5cm}\langle n_R a_L^\dagger  \,a_R \rangle_@ =
\langle a_R^\dagger  \,a_L n_R \rangle_@=\langle n_L a_R^\dagger \,a_L
\rangle_@ =\langle a_L^\dagger  \,a_R n_L \rangle_@=
\frac{(N-1)}{2}\langle a_L^\dagger  \,a_R \rangle_@\,.
\label{nRaLaR}
\eea
From the equation above we can now write the analytical expression for the 
energy as a function of $z$
\be
\label{enz1}
E=-\frac{t}{2}N\left(\frac{\sqrt{1-z^2}+\left(\sqrt{1-z^2}\right)^{N-1}}
{1+\left(\sqrt{1-z^2}\right)^N}\right)+\frac{U}{4}N(N-1)\left(1+\frac{z^2}
{1+\left(\sqrt{1-z^2}\right)^N}\right)\,.
\ee
If $z=0$ we recover the energy in the 
coherent-like state, namely $E=-tN/2+UN(N-1)/4$, while for $z=1$ we get the 
energy in the $NOON$ state, $E=UN(N-1)/2$. 
Eq.~(\ref{enz1}) should be compared with Eq.~(\ref{E_z_phi}) 
(with $\mu=\phi=0$), called now $E_s$, the energy calculated considering 
only a single state, 
$|\psi_L\rangle$ or $|\psi_R\rangle$, separately. For convenience we report 
here $E_s$, the energy of an unbalanced coherent-like state,
\be
\label{enz2}
E_s=-\frac{t}{2}N\sqrt{1-z^2}+\frac{U}{4}N(N-1)\left(1+{z^2}\right)\,.
\ee
Eq.~(\ref{enz2}) has a single minimum in $z=0$ for $U\le U_c$ and two minima 
in $z=z_o$ given in Eq.~(\ref{z}), exhibiting a sort of second order phase 
transition. The concavity at $z=0$, $d^2E_s/dz^2|_{z=0}=n(t+(N-1)U)/2$, 
in fact, is negative for $U<U_c$. The concavity of $E$ at $z=0$, 
instead, is always 
negative for any $U<0$, $d^2E/dz^2|_{z=0}=UN(N-1)/4$. This means that there 
is always at least one finite minimum, i.e. $z_o$ is finite for any $U<0$. 
From Fig.~\ref{fig.E123} we observe a typical picture of a 
first order transition, with a jump in the global minimum at some 
$U_c^*<U_c$, as shown in the left plot of Fig.~\ref{fig.zU}. 
For $U>U_c^*$ therefore, there is a minimum of the energy $E$ given by
\be
z_o\simeq \pm\sqrt{\frac{-2U}{2t+NU}}\,
\ee
while for $U<U_c^*$ the minimum is given by Eq.~(\ref{z}).
In terms of $U_c$, Eq.~(\ref{Uc}), we have, therefore, 
\be
\label{zotot}
z_o\simeq \pm\left\{\ba{lr} \sqrt{\frac{2U}{\left(2U_c-U\right)N}}\,, &\;\;0\ge U>U_c^*\\ 
\sqrt{1-\left(\frac{U_c}{U}\right)^2}\,, &U<U_c^*\ea\right.
\ee
However $U_c^*\rightarrow U_c$ for 
$N\rightarrow \infty$ (see Fig.~\ref{fig.zU}). 
More quantitatively, at leading order, we get 
\be
\label{uc*}
U_c^*\simeq U_c\left(1+\frac{1}{\sqrt{2N}}\right)\,.
\ee
At $U=U_c^*$ there is a small jump in the values of $z_o$, for $N$ 
larger than ten, which is given by 
$\big(1-\frac{2N}{(\sqrt{2N}+1)^2}\big)^{{1}/{2}}-\big(\frac{2(\sqrt{2N}+1)}{N(\sqrt{2N}-1)}\big)^{{1}/{2}} $
and which goes to zero as $(\frac{2}{N})^{{1}/{4}}$, in the large $N$ limit.

What we have learnt is that, only in the limit of 
$N\rightarrow \infty$, which implies 
$E\rightarrow E_s$, the picture of spontaneous symmetry breaking is correct, 
while for finite $N$ there is always a finite imbalance for any attractive 
interaction, but always with a null population imbalance. 

The number fluctuations on each site, 
$\sigma_{L,R}=\langle n_{L,R}^2\rangle_@-\langle n_{L,R}\rangle_@^2$, 
is given by 
\be
\label{sigmaL+}
\sigma=\sigma_L=\sigma_R=\frac{N}{4}+\frac{N(N-1)
\,z^2}{4\left(1+\left(\sqrt{1-z^2}\right)^N\right)}\,,
\ee
which goes from $\sigma=N/4$ in the coherent-like state ($z=0$) 
to $\sigma=N^2/4$ in the $NOON$ state ($z=1$) \cite{sakmann}.

Let us now consider the visibility. Its analytical expression
 can be read out directly from Eq.~(\ref{aLaR}), therefore
\be
\label{alpha+}
\alpha=\left(\frac{\sqrt{1-z^2}+\left(\sqrt{1-z^2}\right)^{N-1}}{1+\left(
\sqrt{1-z^2}\right)^N}\right)\,,
\ee
which is $\alpha=1$ in the coherent-like state while $\alpha\rightarrow 0$ 
for $z\rightarrow 1$, namely going towards the $NOON$ state.

\begin{figure}[h!]
\centering
\includegraphics[width=7cm]{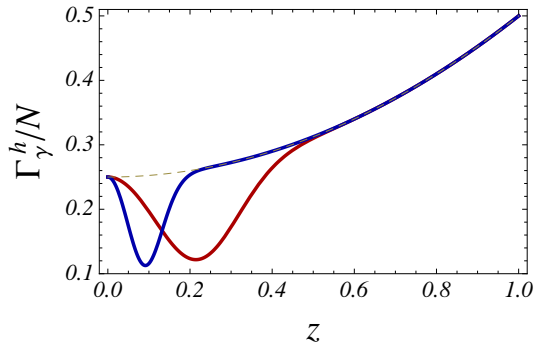}
\caption{(Color online) The contribution $\Gamma_\gamma^h$ to the decay rate, 
rescaled by $N$, as a function of $z$, for $N=300$ (solid blue line, with a 
narrow dip) and $N=50$ (solid red line, with a broad dip). The dashed line is the limit $N\rightarrow \infty$, Eq.~(\ref{gammah_as}).}
\label{fig.gammah}
\end{figure}

Finally we calculate the decay rate in the presence of a weak coupling to an 
external environment, as described before. 
From Eqs.~(\ref{aLaR})-(\ref{nRaLaR}), we get, for $\Gamma_\gamma$, the 
following equation, written in terms of the previous quantities
\be
\label{gamma-}
\Gamma_\gamma=\sigma |c_L-c_R|^2+\Gamma_\gamma^h |c_h|^2+
\alpha\frac{N}{2}\textrm{Re}[(c_L-c_R)c_h^*]
\ee
and, analogously, $\Gamma_\delta=\sigma |c_L-c_R|^2+\Gamma_\gamma^h |c_h|^2-
\alpha\frac{N}{2}\textrm{Re}[(c_L-c_R)c_h^*]$, 
where $\sigma$ is given by Eq.~(\ref{sigmaL+}), $\alpha$ by 
Eq.~(\ref{alpha+}) and 
\be
\Gamma_\gamma^h=\frac{N}{2}-\sigma+\frac{N^2}{4} (1-\alpha^2)\,.
\ee
For $z=0$ ($\alpha=1$, $\sigma=N/4$) we recover the coherent-like result, $\Gamma_\gamma^h=N/4$, while for $z=1$ ($\alpha=0$, $\sigma=N^2/4$) we have $\Gamma_\gamma^h=N/2$. 
For large $N$, the asymptotic form of $\Gamma_\gamma^h$ is
\be
\label{gammah_as}
\Gamma_\gamma^h\xrightarrow[N\rightarrow\infty]{}\frac{N}{4}(1+z^2)
\ee
while, at finite and large $N$, $\Gamma_\gamma^h$ has a minimum at 
$z\simeq 1.6/\sqrt{N}$, where $\Gamma_\gamma^h/N\simeq 0.11$, 
see Fig.~\ref{fig.gammah}.
However, the important point is that $\Gamma_\gamma^h$ does not scale faster 
then $N$. In the presence of a bath coupled differently with the densities 
located on the two sites ($c_L\neq c_R$), instead, the decay rate 
$\Gamma_\gamma$, for $U<U_c^*$, scales as $N^2$ with the number of bosons, 
since the leading term is driven by the number fluctuations, 
$\Gamma_\gamma\simeq\sigma |c_L-c_R|^2$, and $\sigma\sim N^2$ 
for $U<U_c^*$, as one can check from Eq.~(\ref{sigmaL+}), evaluated at $z_o$, 
Eq.(\ref{zotot}).

Before we conclude this section a brief explanation is in order. 
The discontinuity of $z_o$ at $U_c^*$ cannot be directly measured since 
$z_o$ is not an observable, being not equal to the population imbalance, 
which is always $\langle n_L\rangle-\langle n_R\rangle=0$. 
The small jump in $z_o$, however 
could produce a small discontinuity in the visibility or in the variance of 
on-site number of bosons as functions of $U$. 
By exact diagonalization of $H$, instead, those quantities do not exhibit 
any discontinuities. However, the exact results are in a very good agreement 
with our analytical results away from $U_c^*$, while, close to $U_c^*$, 
those quantities, like the visibility, for instance, obtained by exact 
diagonalization, are smooth. 
Nevertheless, their derivatives at $U_c^*$ are large, although not infinite. 
The analytical expression for the energy is also in perfect agreement with the 
exact numerical result $E_{ex}$, see the last plot in Fig. \ref{fig.E123} and 
Fig. \ref{fig.ener}. 
It behaves as $\sim UN^2/4-tN/2$ and 
$\sim UN^2/2-tN$ for, respectively, very small and very large interactions. 
The point of crossover is identified by looking at the mimimum of the 
second derivative of $E$ with respect to $U$ (see the right plot of Fig. 
\ref{fig.zU}) which coincides with $U_c^*$, the point where 
$(\textrm{min}(E)-\textrm{min}(E_s))\simeq 0$ (compare 
the right plot and the central plot of Fig. \ref{fig.zU}). We have 
checked this result also for different values of $N$. Therefore, this value 
of interaction, in the presence of a finite number of bosons, can be promoted 
as the crossover point, which characterizes the loss of coherence and, as 
we will see in the 
next section, is also the point where the quantum entanglement reaches its 
maximum. Finally, for $U<U_c^*$, as explained in the following paragraph, 
the cat state becomes extremely fragile under a small offset between the 
two on-site energies, i.e. $\mu\neq 0$. This fragility is also revealed, as we 
have seen before, by 
the decay rate in Eq.~(\ref{gamma-}), when the two sites are coupled 
differently to a source of noise induced by an external bath, 
i.e. $c_L\neq c_R$, since, in this case, the 
instability is driven by the large number fluctuations which goes like $N^2$ 
for $U$ more negative than $U_c^*$.


\subsection{Breakdown of the cat state}

The fragility of the Schr\"odinger cat state can be shown not only by 
analyzing the decay time $\tau_\gamma=1/\Gamma_\gamma$, which 
goes like $1/N^2$, for $U<U_c^*$ and $c_L\neq c_R$, 
but also by energetic arguments. 
As one can see from Fig.~\ref{fig.zU} (central plot) the energy difference 
$\Delta E_o$ between the minimum of $E$, 
calculated in the cat state, and the minimum of $E_s$ calculated for a 
coherent-like state with finite imbalance $z$, is always positive and becomes 
almost zero for $U<U_c^*$. Actually, the positions of the global minima of the 
two energies almost coincide, below that critical interaction, 
see Fig.~\ref{fig.E123}. However, the energy obtained using the cat state is 
unaffected by a finite value of the chemical potential difference $\mu$ 
between the two sites, unlike the energy of a coherent state. The latter is 
modified by an additional term $\mu N z$. Therefore, for large $N$ and 
$U<U_c^*$ an almost infinitesimal value of $\mu$ makes the cat state, 
superposition of two unbalanced coherent-like states, 
energetically unfavourable with respect to a single coherent-like state.
In this case, the number fluctuations and the visibility become simply those 
in Eqs.~(\ref{sigmaC}),~(\ref{alphaC}), or, in terms of $z$,
\bea
\label{sigmaz}
&&\sigma=\frac{N}{4}(1-z^2)\,,\\
&&\alpha=\sqrt{1-z^2}\,.
\eea
Notice that $\sigma$ drops the $N^2$ dependence and becomes linear in the number of bosons. 
Finally, let us consider the decay rate. If $\mu <0$ the cat state collapses 
to $|\psi_L\rangle$ and 
the decay rate $\Gamma_\gamma$, in terms of $z$, is given by
\be
\Gamma_\gamma=\frac{N}{4}\left\{|c_L-c_R|^2(1-z^2)+|c_h|^2(1-z)^2+
2\sqrt{1-z^2}\,\textrm{Re}[(c_L(1+z)-c_R(1-z))c_h^*]\right\}\,.\\
\ee
For $\Gamma_\delta$ one has to interchange $c_L$ with $c_R$ ($c_L\leftrightarrow c_R$) and change the sign of $z$ ($z\rightarrow -z$). 
If $\mu >0$, the ground state is, instead, $|\psi_R\rangle$, therefore, the decay rates are the sames as before, providing that $z\rightarrow -z$.

\section{Entanglement for $U\le 0$}

\subsection{Entropy}
\begin{figure}[h!]
\centering
\includegraphics[width=7cm]{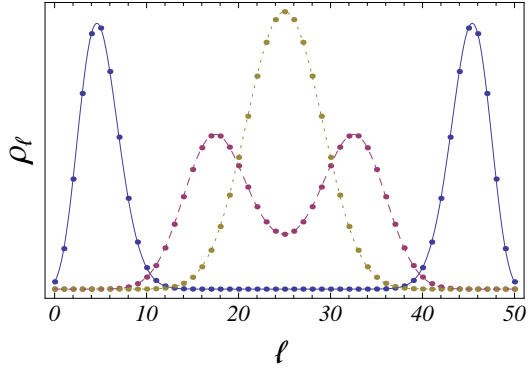}
\caption{(Color online) Reduced density matrix for $N=50$ and $z=0.1$ (yellow dotted line), $z=0.3$ (red dashed line), $z=0.8$ (blue solid line). 
The lines are guides for the eye.}
\label{fig.rho}
\end{figure}

After tracing over one site, for instance, the right one, we get
\be
\label{rho_sb}
\hat\rho=\frac{1}{\langle\psi_@|\psi_@\rangle}\textrm{Tr}_R\left(|\psi_@\rangle \langle\psi_@|\right)=
\frac{N!}{\langle\psi_@|\psi_@\rangle}\left(\hat\rho^{LL}+\hat\rho^{RR}
+\hat\rho^{LR}+\rho^{RL}\right)
\ee
where $\hat\rho^{ss'}=\frac{1}{N!}\textrm{Tr}_R\left(|\psi_s\rangle \langle\psi_s'|\right)$, are diagonal matrices with finite elements given by
\bea
&&\rho^{o}_\ell\equiv
\rho^{LL}_{\ell}=\rho^{RR}_{N-\ell}=\Big(\ba{c} N\\\ell\ea \Big)
|\xi^>|^{2\ell}|\xi^<|^{2(N-\ell)}\,,\\
&&\rho^{LR}_{\ell}=\rho^{RL}_{N-\ell}=\Big(\ba{c} N\\\ell\ea \Big)
(\xi^{>*}\xi^{<})^{\ell}(\xi^{<*}\xi^{>})^{(N-\ell)}\,,
\eea
where $|\xi^>|^2+|\xi^<|^2=1$. 
Eq.~(\ref{rho_sb}) is therefore a diagonal matrix, 
$\rho_{\ell\ell'}=\delta_{\ell\ell'}\rho_\ell$, with
\be
\rho_\ell=\frac{1}{2\left(1+\big(\textrm{Re}[\xi^{>*}\xi^<]\big)^N\right)}
\left(\rho^{o}_\ell+\rho^{o}_{N-\ell}+
\Big(\ba{c} N\\\ell\ea\Big)\textrm{Re}\left[\big(\xi^{>*}\xi^<\big)^N\right]\right).
\ee
As already said, at the ground state $\phi=2n\pi$, therefore $\cos\phi=\cos N\phi=1$, and remembering that $|\xi^{>,<}|^2=(1\pm z)/2$, we have simply
\be
\label{rho-}
\rho_\ell=\frac{1}{2\left(1+\big(\sqrt{1-z^2}\big)^N\right)}
\left(\rho^{o}_\ell+\rho^{o}_{N-\ell}+
\frac{1}{2^{N-1}}\Big(\ba{c} N\\\ell\ea\Big)\left(\sqrt{1-z^2}\right)^N
\right).
\ee
where $\rho^o_\ell$, in terms of $z$, is given by 
$\rho^{o}_\ell=\frac{1}{2^N}\Big(\ba{c} N\\\ell\ea \Big)(1+z)^{\ell}
(1-z)^{(N-\ell)}$. It is clear, also by looking at Fig.~{\ref{fig.rho}}, that 
the profile of the reduced density matrix
has a bimodal shape, peaked at the values $N(1+z)/2$ and $N(1-z)/2$, with 
an additional interference term. When $z$ is small the two peaks merge 
together, while for $z$ close to $1$ the two peaks are far apart. \\
For $z=0$, in fact, we recover the coherent-like reduced density matrix
\be
\rho_\ell=\frac{1}{2^N}\Big(\ba{c} N\\\ell\ea \Big)\,,
\ee
while for $z=1$, namely in the $NOON$ state, we have
\be
\label{rhonoon}
\rho_\ell=\frac{1}{2}\left(\delta_{\ell,0}+\delta_{\ell,N}\right)\,.
\ee
Now we can calculate the entanglement von Neumann entropy, whose expression is here reported,
\be
\label{S-}
S=-\sum_{\ell=0}^N \rho_\ell \log_2{\rho_\ell}\,.
\ee
For $z=0$ and for large $N$ the reduced 
density matrix approaches a gaussian distribution, $\rho_\ell\simeq 
\frac{1}{\sqrt{\pi N/2}}\exp\left[-\frac{(\ell-N/2)^2}{N/2}\right]$, 
therefore, the asymptotic behavior for the entanglement entropy in the 
coherent-like state is given by Eq.~(\ref{Sco}), namely
\be
\label{Sz0}
S_{o}=\frac{1}{2}\log_2\left(\frac{\pi e}{2}N\right)\,.
\ee
For $z=1$, in the $NOON$ state, instead, 
since the reduce density matrix is given by Eq.~(\ref{rhonoon}), and then the entropy is simply 
\be
S_{1}=1\,.
\ee
For finite $z$, in the large $N$ limit, we can approximate the reduced density matrix as $\rho_\ell\simeq \frac{1}{2}(\rho_\ell^o+\rho_{N-\ell}^o)$, i.e. as the sum of two gaussians, therefore, if $z$ is large enough that the two gaussians are well separated, 
the entropy has the following asymptotic form
\be
\label{Sapz}
S_z\simeq \frac{1}{2}\log_2\left(2{\pi e} N(1-z^2)\right)\,,
\ee
where, of course, $z^2<1-2/\pi e N$ in order not to go below the $NOON$ entropy 
or even to diverge.  
\begin{figure}[h!]
\centering
\includegraphics[width=7cm]{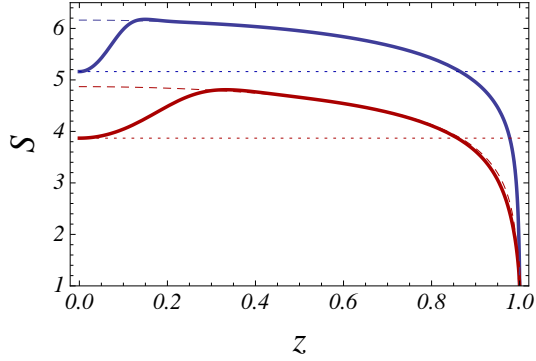}
\caption{(Color online) Entanglement entropy, $S$, in the cat state, 
as a function of $z$, 
for $N=300$ (upper solid line) and $N=50$ (lower solid line).
The dashed lines are the asymptotic behaviors given by Eq.~(\ref{Sapz}), 
for $N=300$ (upper dashed line) and $N=50$ (lower dashed line). 
which match the exact entropy for $z>2.6/\sqrt{N}$. 
The straight dotted lines are the values of $S$ for $z=0$, 
given by Eq.~(\ref{Sz0}), i.e. $S_o$, the entropy of the coherent state, 
for $N=300$ (upper dotted line) and $N=50$ (lower dotted line).}
\label{fig.Sz}
\end{figure}
Approaching $z\rightarrow 0$ the 
two separated gaussians of the reduced density matrix, see Fig.~\ref{fig.rho}, 
which gives the entropy in Eq.~(\ref{Sapz}), start to see each other when  
the distance between the two peaks, which is $Nz$, 
becomes smaller than a certain number, $n_\sigma$, 
of standard deviations, $\sigma_g=\sqrt{N(1-z^2)/4}$, of each gaussian. 
Namely, the 
entropy reaches its maximum and then starts to decrease approaching $z=0$, 
when $z$ fulfils
\be
Nz \simeq 2\,\sigma_g n_\sigma\,,
\ee
whose solution is given by
\be
z_{S_{max}}\simeq \frac{n_\sigma}{\sqrt{n_\sigma^2+N}}\,.
\ee
From Fig.~\ref{fig.Sz}, and further checks, in fact,
we have found that the number of standard deviations,
consistent with such a description, is $n_\sigma \approx 2.6$, therefore the
maximum value for the entropy is reached at $z_{S_{max}}\approx 2.6/\sqrt{N}$,
corresponding to having a distance between the two peaks of the reduced
density matrix approximately equal to five sigmas.
If $z$ were given by Eq.~(\ref{z}), the entropy would reach its 
maximum at
$
U_{S_{max}}\simeq 
U_c\sqrt{\frac{N+n_\sigma^2}{N}} \simeq 
U_c\left(1+\frac{n_\sigma^2}{2 N}\right)$.
However, since $z_o|_{U_c^{*}-\varepsilon}<z_{S_{max}}<z_o|_{U_c^{*}+\varepsilon}$, then the maximum entropy occurs at
\be
\label{Umax}
U_{S_{max}}=U_c^*\,,
\ee
where $U_c^*$ is given by Eq.~(\ref{uc*}).
For large $N$, $z_{S_{max}}$ goes to zero, and, as a result, 
the asymptotic value of the maximum entropy can be written simply 
substituting $z$ with $0$ in Eq.~(\ref{Sapz}), getting
\be
S_{max}\simeq \frac{1}{2}\log_2\left(2{\pi e} N\right)=\,S_{o}+1\,,
\ee
which does not depend on the interaction strength. In conclusion, the entropy 
grows up to $1$ above its value at $U=0$, Eq.~(\ref{Sz0}), within a 
short range of interaction close to the value $U_c^*$, 
Eq.~(\ref{Umax}). Strikingly, from Eq.~(\ref{Sapz}), we have that the entropy 
exceeds its coherent value ($S> S_o$) for $0< z \lesssim \sqrt{3}/2$. 
In terms of the interaction, this means that 
\be
S> S_o\,, \;\; \textrm{for} \;\;\,2\,U_c\lesssim U< 0.
\ee

\subsection{Fisher information}
From the definition of Fisher information, Eq.~(\ref{F2}), 
and after verifying that
\be
\langle n_Ln_R\rangle_@-\langle n_L\rangle_@\langle n_R\rangle_@=-\sigma
\ee
where $\sigma$ is given by Eq.~(\ref{sigmaL+}), 
we find the following remarkable exact result 
\be
\label{Fsigma}
F=\frac{4\,\sigma}{N^2}\,,
\ee 
valid along all the attractive regime, from $z=0$ 
($U=0$, coherent-like state), where $\sigma=N/4$ and $F=1/N$, to $z=1$ 
($U\rightarrow -\infty$, $NOON$ state), where $\sigma=N^2/4$ and $F=1$. 
Even if the system collapses into an unbalanced coherent-like state, 
Eq.~(\ref{Fsigma}) is still valid, providing that $\sigma$ is given by 
Eq.~(\ref{sigmaz}). Surprisingly, Eq.~(\ref{Fsigma}) is the same relation 
as that obtained also for repulsive interaction, Eq.~(\ref{Fsigma+}).

\section{Summary of main results with plots}
\begin{figure}[h!]
\centering
\includegraphics[width=6cm]{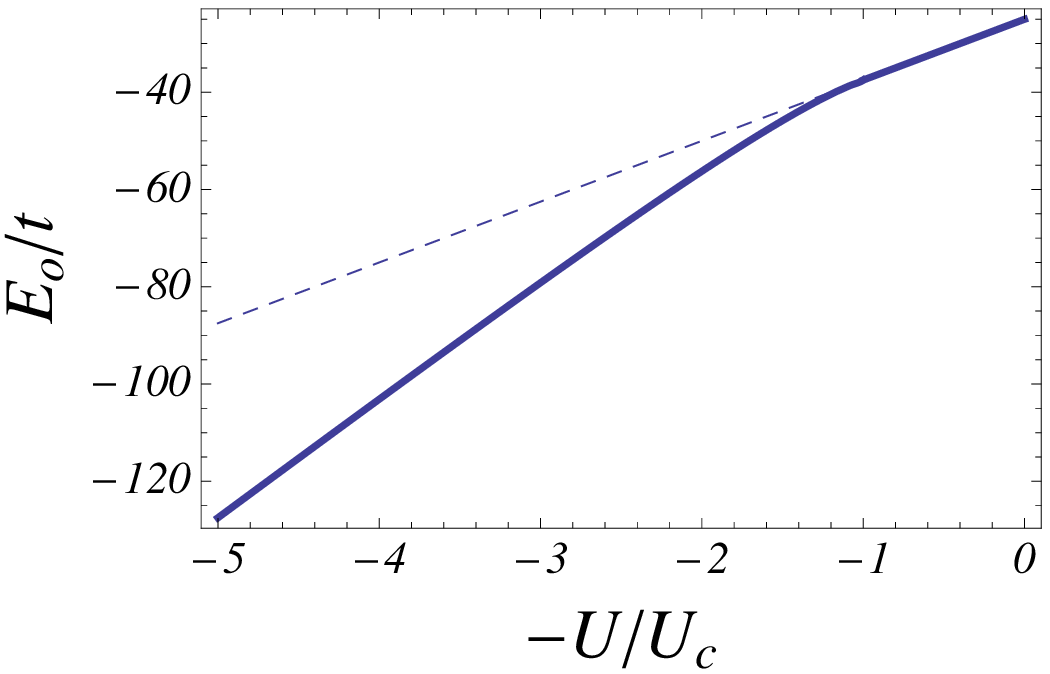}\hspace{0.3cm}
\includegraphics[width=6cm]{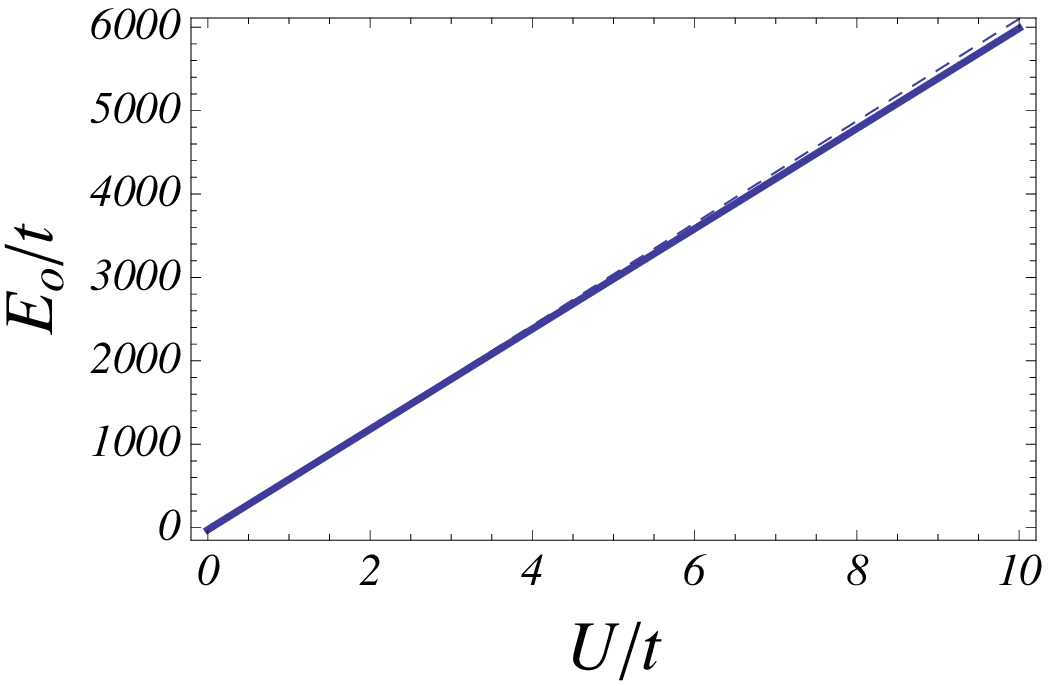}
\caption{(Color online) Ground state energy $E_o=\textrm{min}(E)$ as a function of $U$ for $N=50$, for attractive (left plot) and repulsive (right plot) interactions. Left plot (attractive interaction): 
the energy is described by Eq.~(\ref{enz1}) at $z_o$ given by 
Eq.~(\ref{zotot}). Right plot (repulsive interaction): 
the energy is described by Eq.~(\ref{enmin}). 
In both the plots, the dashed line is 
the coherent energy, $E_C=-t/2+UN(N-1)/4$, showing that a 
finite imbalance lowers the ground state energy.
The ground state energy obtained by exact diagonalization, $E_{ex}$, 
coincides, on the scale of the plots, with the analytical result $E_o$ 
(solid line), in both the regimes.}
\label{fig.ener}
\end{figure}

\begin{figure}[h!]
\centering
\includegraphics[width=6cm]{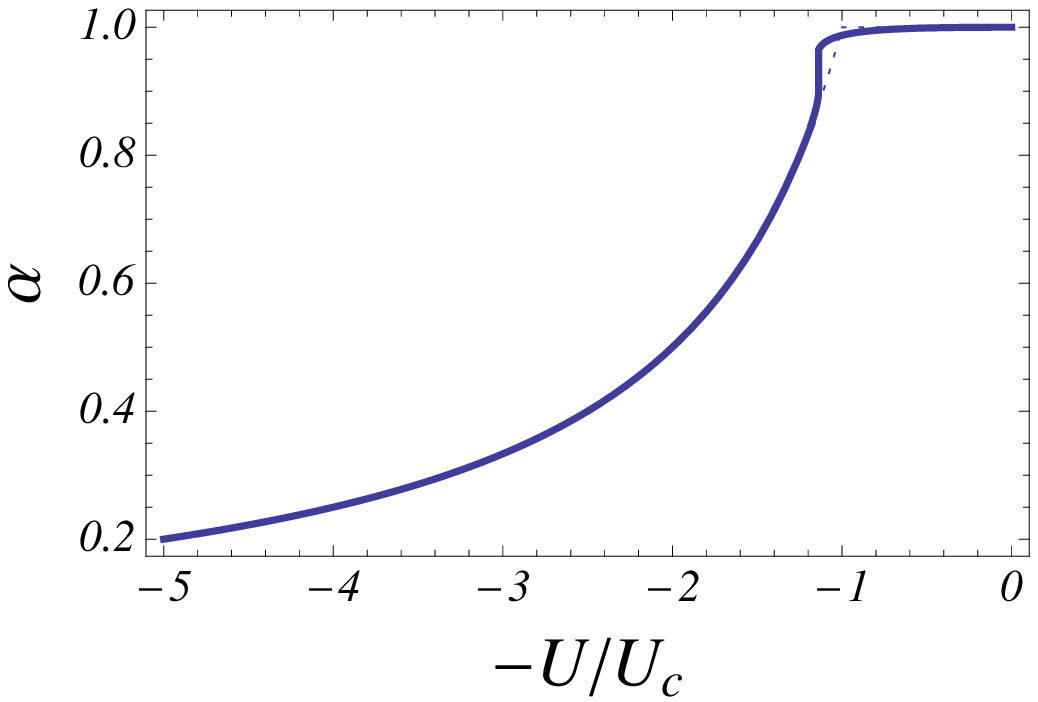}\hspace{0.3cm}
\includegraphics[width=6cm]{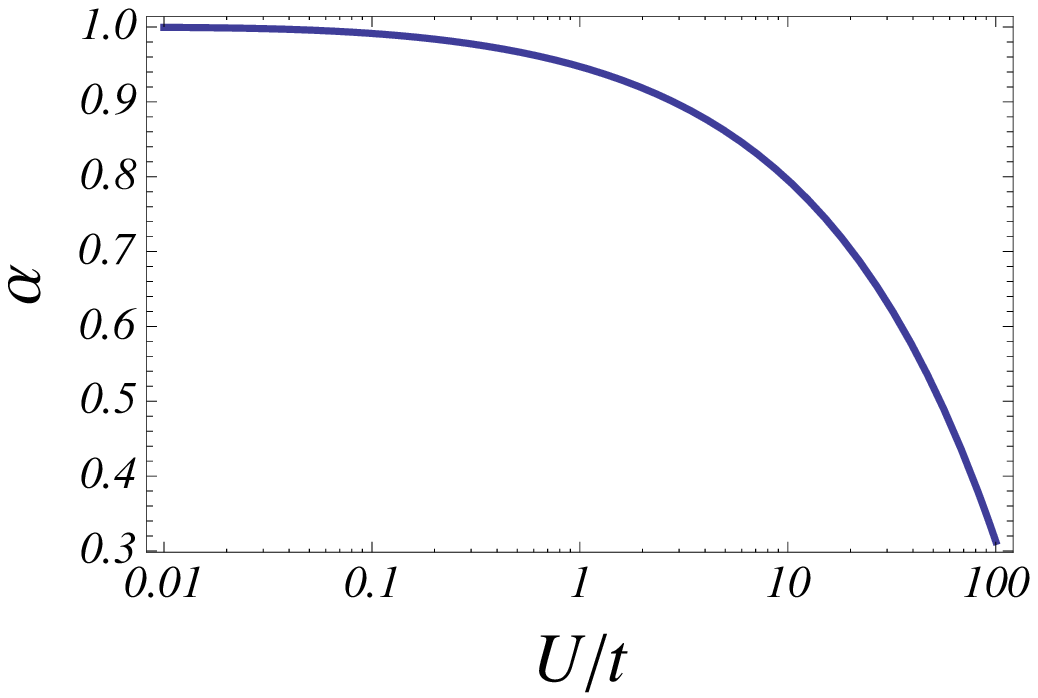}
\caption{(Color online) Visibility $\alpha$ as a function of $U$ for $N=50$, for 
attractive (left plot) and repulsive (right plot) interactions. 
Left plot (attractive interaction):
the visibility as in Eq.~(\ref{alpha+}), calculated at $z_o$ given by
Eq.~(\ref{zotot}). 
The dotted line is the limit $N\rightarrow\infty$.
Right plot (repulsive interaction):
the visibility described by Eq.~(\ref{alpha}).}
\label{fig.visib}
\end{figure}

\begin{figure}[h!]
\centering
\includegraphics[width=6cm]{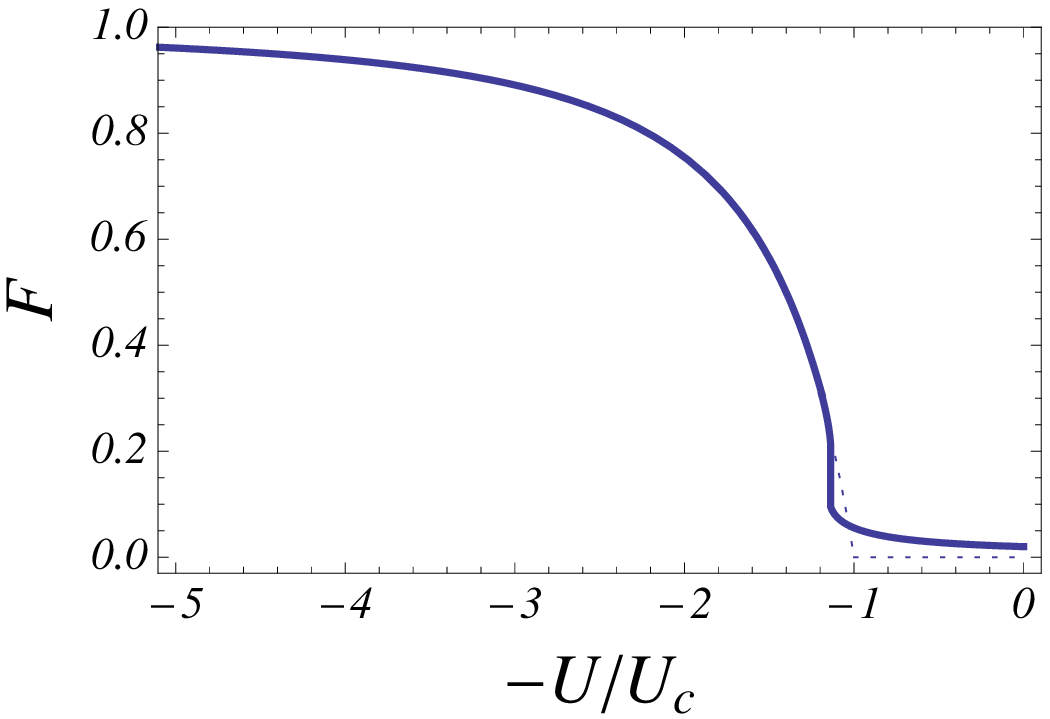}\hspace{0.3cm}
\includegraphics[width=6cm]{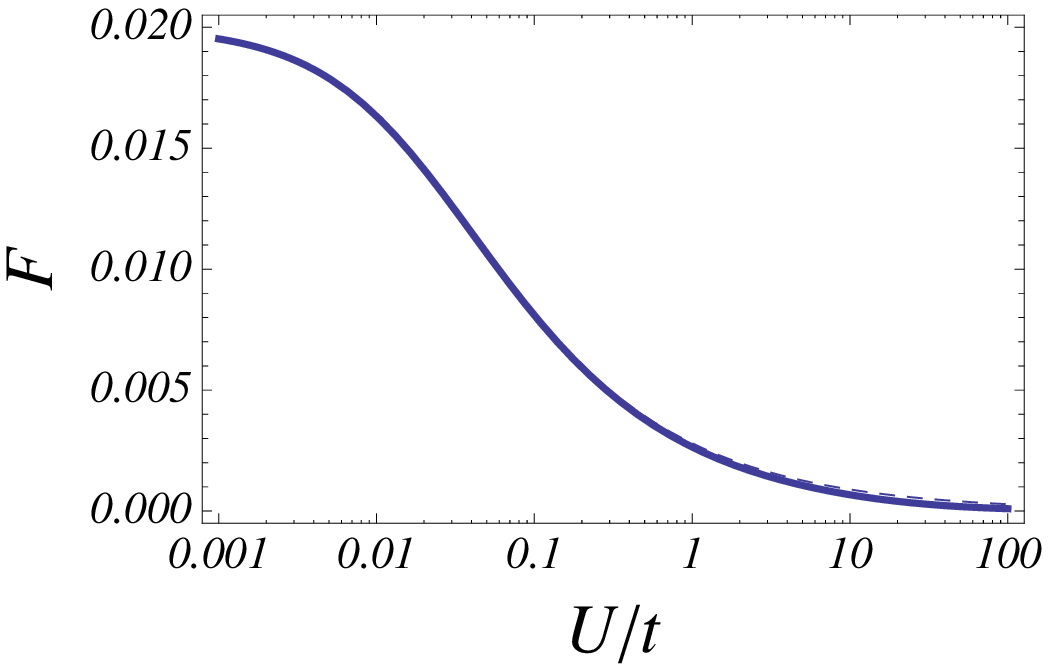}
\caption{(Color online) Fisher information $F$ as a function of $U$, for $N=50$, for 
attractive (left plot) and repulsive (right plot) interactions. 
The Fisher information is described by $F=4\sigma/N^2$ where $\sigma$ 
is the variance of the number of particles. 
Left plot (attractive interaction): $\sigma$ is given by Eq.~(\ref{sigmaL+}), 
calculated at $z_o$ in Eq.~(\ref{zotot}). 
The dotted line is the limit $N\rightarrow\infty$.
Right plot (repulsive interaction): 
$F$ given by Eq.~(\ref{F}), calculated from full correlators, Eqs.~(\ref{PerD}), (\ref{PerI}), with parametrizations Eq.~(\ref{param1}), (\ref{param2}), 
at $\omega=\omega_o$, Eq.~(\ref{wmin}). The dashed line, which almost coincides with the solid one, is simply given
by Eq.~(\ref{Fomega}) at $\omega_o$.}
\label{fig.F}
\end{figure}

\begin{figure}[h!]
\centering
\includegraphics[width=6cm]{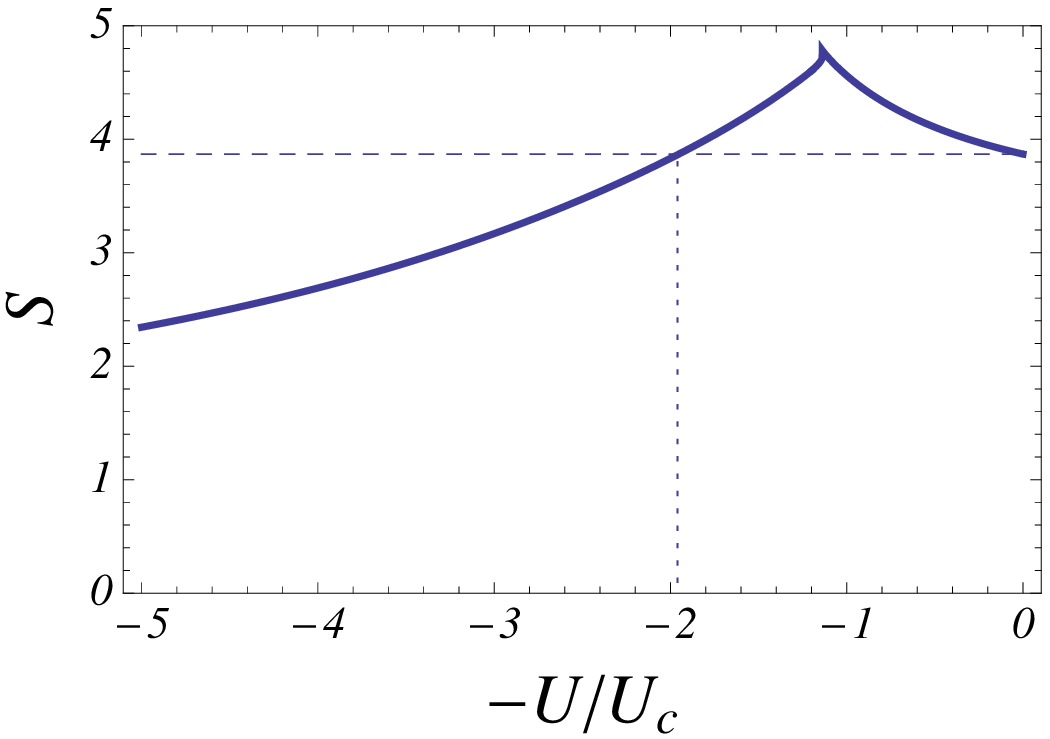}\hspace{0.3cm}
\includegraphics[width=6cm]{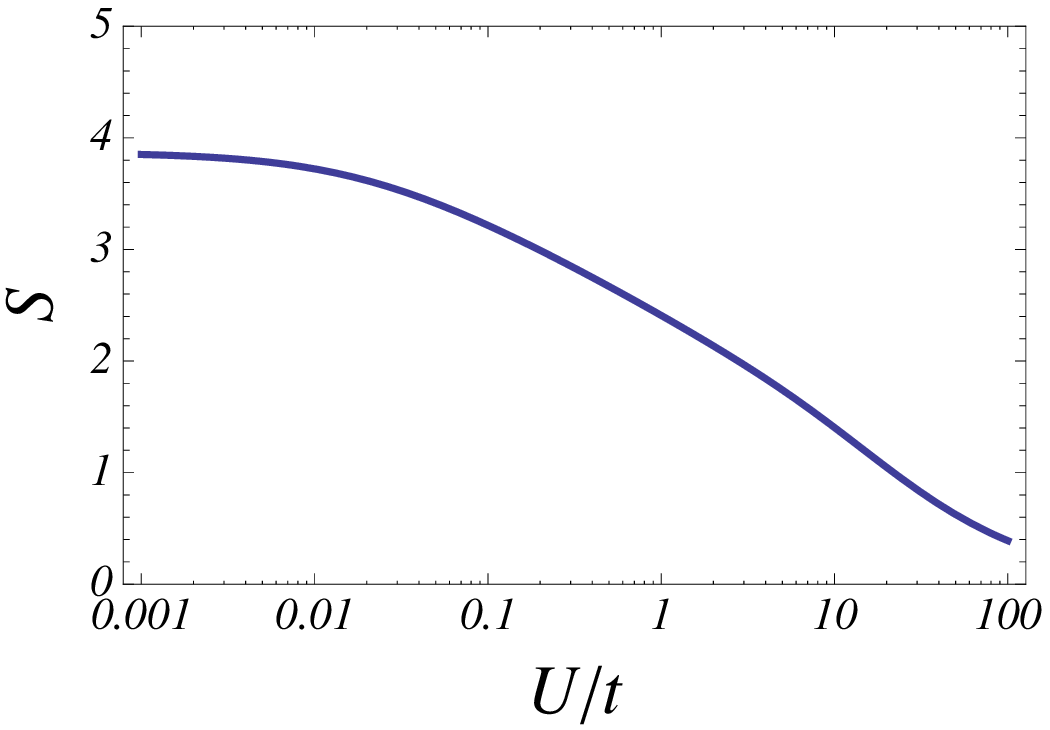}
\caption{(Color online) Entanglement entropy $S$ as a function of $U$, for $N=50$, for 
attractive (left plot) and repulsive (right plot) interactions. Left plot 
(attractive interaction): $S$ is given by Eq.~(\ref{S-}), 
with $\rho_\ell$ obtained from Eq.~(\ref{rho-}), 
calculated at $z_o$ in Eq.~(\ref{zotot}). The dashed line corresponds to 
$S_o$, Eq.~(\ref{Sz0}). $S$ exceeds $S_o$ for $2U_c\lesssim U<0$.
Right plot (repulsive interaction): 
entropy calculated through Eqs.~(\ref{rho1}), (\ref{rho2}), with $n=k=N/2$ and 
Eqs.~(\ref{param1}), (\ref{param2}), at $\omega_o$, Eq.~(\ref{wmin}).}
\label{fig.S}
\end{figure}

In this section we show some plots of several quantities: energy 
(Fig.~\ref{fig.ener}), coherence visibility (Fig.~\ref{fig.visib}), 
Fisher information (Fig.~\ref{fig.F}), and entanglement entropy 
(Fig.~\ref{fig.S}), all as functions 
of the interaction strength, both in the attractive and repulsive regimes.
We used $N=50$ in order to show features visible only for not too large number of particles. 

\section{Conclusions}
We have studied, in the repulsive regime, the interpolation from a spatial 
separated Fock state, ground 
state in the limit of strong repulsive interaction of the Bose-Hubbard 
Hamiltonian, to a delocalized coherent state, ground state for free bosons. 
We have used as many-body wavefunction a simplified permanent state which 
allows us to write all the physical quantities we are 
interested in in terms of 
polynomials of the single particle overlap parameter which can be determined 
variationally. Although not exact, our single particle approach provides 
a quite transparent and intuitive description of the crossover between the two 
exact limits. 
We have calculated the energy, the charge fluctuations, the decay rate, 
the coherence visibility, the entanglement entropy and the Fisher information 
with such a trial wavefunction.
We have derived a non-perturbative, simple analytical expression for the 
ground state energy in the large $N$ limit.
We have shown that the charge fluctuation, null in the 
Fock limit while proportional to $N$ for zero interaction, in the 
presence of finite interaction scales as $N^{1/2}$ with the number of bosons. 
Moreover we have shown that the decay rate, in the presence of a weak 
coupling with an external environment, scales as $N^{3/2}$ in the 
intermediate case, i.e. finite $U$, and that it is proportional to the 
inverse of the Fisher information. 
In this regard, the overlap of single bosons can be considered as the key 
parameter which contains the quantum information. Moreover we have shown 
that, for $U\ll N$, the Fisher information $F$ and the on-site number 
fluctuations expressed by the variance $\sigma$ are related by the equation 
$F=4\sigma /N^2$. 
In the attractive regime we have calculated again the energy, the  
variance of bosonic numbers $\sigma$, the decay rate, the visibility,  
the Fisher information and the entanglement entropy by using a symmetric 
superposition of imbalanced coherent-like states, namely a Schr\"odinger cat 
state. We have shown that this state, in the absence of a local offset between 
the two on-site energies, has always a lower energy with respect to a single coherent-like 
state which usually is supposed to describe well the low interaction regime. 
As a result, the symmetry breaking for finite number of bosons does not occur, 
and the corresponding critical interaction $U_c$, 
shifted toward more negative $U$, 
i.e. $U_c^*\simeq U_c(1+\frac{1}{\sqrt{2N}})$, 
is actually a crossover point separating two different regimes, the 
coherent regime and the incoherent one; in the latter regime the cat state 
becomes extremely fragile. In the presence of an external bath, if the 
environment is coupled differently to the two local densities the 
relaxation time is very short, inversely proportional to the number 
fluctuations which go like $N^2$ for $U<U_c^*$. 
Remarkably, we have found that 
the equation for the Fisher information, derived in the repulsive regime, 
i.e. $F=4\sigma /N^2$, is exactly valid also for the whole range of negative 
interactions. 
Finally, we have calculated the entanglement entropy and found that, close to 
the crossover point, $U_c^*$, the entropy reaches its maximum value, 
which, in the large $N$ limit, is equal to the entropy at $U=0$, 
increased by one (notice that $1$ is the entropy in the $NOON$ state). 
Another important and final result, valid for large $N$, is that the 
entanglement entropy exceeds its coherent value (at $U=0$) 
for $2U_c\lesssim U< 0$. 

\begin{acknowledgments}
I thank F. Benatti, R. Floreanini, M. Fabrizio, G. Mazzarella, 
L. Salasnich, F. Toigo and A. Trombettoni for useful discussions and 
acknowledge financial support by the 
Department of Physics -Miramare-, University of Trieste, 
during the first stage of the work, and hospitality by SISSA, Trieste. 
I dedicate this work to Sasha Gogolin, a dear friend and great scientist.
\end{acknowledgments}

\end{document}